\documentclass{osa-article}

\journal{osac}
\articletype{Research Article}
\begin{document}

\title{A novel two-mode squeezed light based on double-pump phase-matching}

\author{Xuan-Jian He,\authormark{1} Jun Jia,\authormark{1} Gao-Feng Jiao,\authormark{1} L. Q. Chen,\authormark{1,3,4} Weiping Zhang,\authormark{2,3} and Chun-Hua Yuan\authormark{1,3,*}}

\address{\authormark{1}State Key Laboratory of Precision Spectroscopy, Quantum Institute for Light and Atoms, Department of Physics, East China Normal University, Shanghai 200062, China\\
\authormark{2}School of Physics and Astronomy, and Tsung-Dao Lee Institute, Shanghai Jiao Tong University, Shanghai 200240, China\\
\authormark{3}Collaborative Innovation Center of Extreme Optics, Shanxi
University, Taiyuan, Shanxi 030006, China\\
\authormark{4}lqchen@phy.ecnu.edu.cn}

\email{\authormark{*}chyuan@phy.ecnu.edu.cn} %% email address is required

\begin{abstract}
A novel two-mode non-degenerate squeezed light is generated based on a
four-wave mixing (4WM) process driven by two pump fields crossing at a small
angle. By exchanging the roles of the pump beams and the probe and conjugate
beams, we have demonstrated the frequency-degenerate two-mode squeezed light
with separated spatial patterns. Different from a 4WM process driven by one
pump field, the refractive index of the corresponding probe field $n_{p}$ can
be converted to a value that is greater than $1$ or less than $1$ by an angle
adjustment. In the new region with $n_{p}<1$, the bandwidth of the gain is
relatively large due to the slow change in the refractive index with the
two-photon detuning. As the bandwidth is important for the practical
application of a quantum memory, the wide-bandwidth intensity-squeezed light
fields provide new prospects for quantum memories.
\end{abstract}

%%%%%%%%%%%%%%%%%%%%%%%%%%  body  %%%%%%%%%%%%%%%%%%%%%%%%%%

\section{Introduction}

Memory for quantum states of light is a necessary component for any future
quantum optical computer \cite{Simon10}. In order to extend the storage
procedure to squeezed states , we need squeezed light that is resonant to the
atomic medium we are using for storage, namely $^{87}$Rb or $^{85}$Rb. The
generation of squezeed light at atomic wavelengths have been obtained on the
rubidium D1 line \cite{Tanimura,Hetet} and D2 line \cite{Polzik,Marin,Burks09}%
. The squeezed vacuum state on the rubidium D1 line has been stored
\cite{Honda, Appel}. Furthermore, the bandwidth is important for the practical
application of a quantum memory \cite{Guo19}. The generated wide-bandwidth
intensity-squeezed light fields at atomic wavelengths provides new prospects
for a quantum memory. Therefore, it is worth initiating a study on how to
generate a two-mode squeezed state of wide-bandwidth, especially frequency
degenerate two-mode squeezed state of wide-bandwidth.

The first experimental demonstration of squeezed states of light by Slusher
\emph{et al.} \cite{Slusher} was based on four-wave mixing (4WM) in sodium
vapor. Since then, many techniques for producing different types of squeezing
have been explored, each with its own advantages and limitations for
particular applications \cite{Schnabel}. Nondegenerate 4WM in a
double-$\Lambda$ scheme \cite{Hemmer} was identified as a possible scheme to
generate a squeezed state or squeezed twin beams, as described in Refs.
\cite{Lukin,Zibrov,Balic,Kolchin,Thompson,Wal,McCormick07,McCormick08}.

\begin{figure}[ptbh]
\centerline{\includegraphics[scale=1,angle=0]{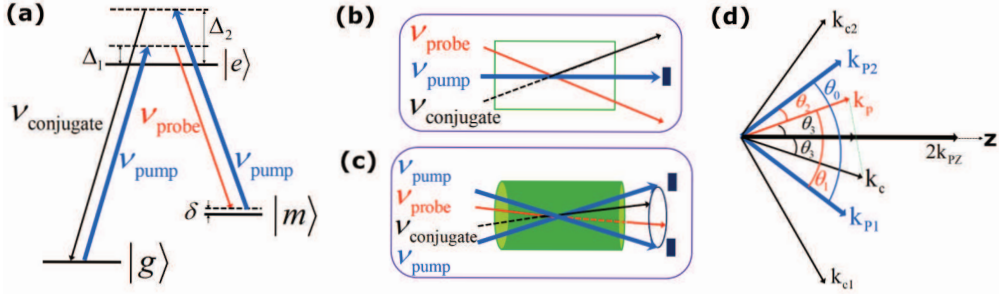}} \caption{(a)
Double-$\Lambda$ configuration for the generation of 4WM with two
frequency-degenerate pump fields for (b) copropagation, and (c) non-collinear
propagation. (d) The angle between pump fields $P_{1}$ and $P_{2}$ is
$\theta_{0}$. The angles between the probe field $p$ and the pump fields
$P_{1}$ and $P_{2}$ are $\theta_{1}$\ and $\theta_{2}$, respectively. The
wavevector $\mathbf{k}_{PZ}$ is the projection of the pump field $P_{1}$ or
$P_{2}$ onto the $z$-axis. When only one pump field $P_{1}$ or $P_{2}$ exists,
the conjugate field $c_{1}$ or $c_{2}$ is generated under the respective phase
matching condition. When two pump fields $P_{1}$ and $P_{2}$ exist at the same
time, a new conjugate field $c$ is generated due to the new phase matching
condition. The probe field $p$ and the conjugate field $c$ are at an angle of
$\theta_{3}$ relative to the $z$-axis. }%
\label{fig1}%
\end{figure}

The generated twin beams by the 4WM process in atomic system with higher
squeezing degree were firstly realized by McCormick \emph{et al.}
\cite{McCormick07,McCormick08} based on degenerate pump fields, as shown in
Fig.~\ref{fig1}(a). A single linearly polarized pump beam, $\nu_{\mathrm{pump}%
}$, is crossed at a small angle with an orthogonally polarized, much weaker
probe beam, $\nu_{\mathrm{probe}}$. The 4WM process amplifies the probe and
generates a quantum-correlated conjugate beam, $\nu_{\mathrm{conjugate}}$, on
the other side of the pump (at a higher frequency), as shown in
Fig.~\ref{fig1}(b). In this case, a pair of photons of the (single) pump is
transformed, via the 4WM process, into a photon in the probe beam and a photon
in the conjugate beam. By modulating the involved ground (excited) state with
one (two) laser beam (beams), the gain and squeezing degree can be enhanced
\cite{Zhang15,Zhang17}. The best initial results for two-mode
intensity-difference squeezing at low frequencies seem to be $\approx1.5$ kHz
\cite{Liu11} to the recently reported $\approx$700 Hz \cite{Ma18} or even
$\approx$10 Hz \cite{Wu19}. The generated entanglement between the probe and
conjugate beams can realize quantum imaging \cite{Boyer08,Boyer082}. The
cascaded 4WM can generate the quantum correlated triple beams
\cite{Qin14,Qin15} and can also be used to realized SU(1,1) interferometers
for highly sensitive phase measurements \cite{Hudelist,Du18}. This 4WM process
also supports many spatial modes, making it possible to amplify complex
two-dimensional spatial patterns \cite{Corzo12,Embrey15,Wang17,Cao17}.

Recently, as shown in Fig.~\ref{fig1}(c), a new 4WM process driven by two pump
fields of the same frequency crossing at a small angle was realized
\cite{Jia,Knutson18}. Instead of two superimposed rings centered around the
pump beams, we find that the output field is satisfied with a two-pump forward
phase matching geometry and is two-beam excited conical emission
\cite{Kauranen}. That is, the light is emitted on the surface of a circular
cone centered on the bisector of the two pump beams. In this paper, we further
implement frequency-degenerate two-mode squeezed light based on a 4WM driven
by two pump fields crossing at a small angle through an optical phase locked
loop (OPLL) \cite{Fox} and give theoretical explanations. By analyzing the
gain, we find that the phase matching condition can be achieved under the
conditions of $n_{p}>1$ and $n_{p}<1$ by an angle adjustment. The theoretical
range of angles for achieving different regions is given. In the new region
with $n_{p}<1$, the bandwidth of the gain is relatively large due to the slow
change in the refractive index with the two-photon detuning, which is
advantageous for realizing wide-bandwidth intensity-squeezed light.

\begin{figure}[ptbh]
\centerline{\includegraphics[scale=0.6,angle=0]{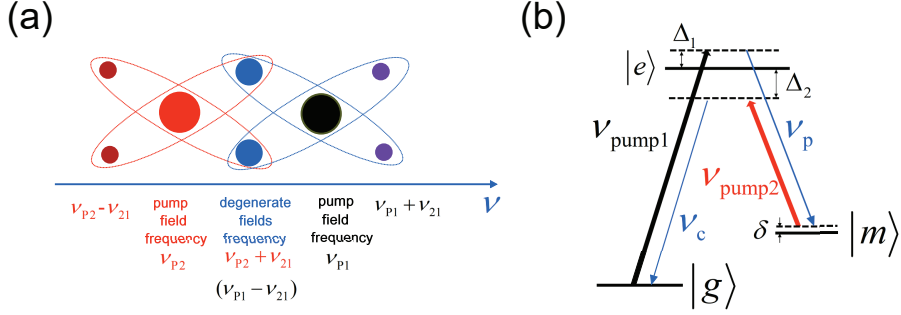}} \caption{(a)
Transverse section of the exiting optical port after exchanging the roles of
the pump beams and the probe and conjugate beams directly. The beams of
frequencies $\nu_{P2}-\nu_{21}$ and $\nu_{P1}+\nu_{21}$ are the unexpected
nonlinear processes. (b) Schematic diagram of frequency-degenerate squeezed
light with $\Delta_{1}$ $(=\omega_{P1}-\omega_{31})$ and $\Delta_{2}$
$(=\omega_{P2}+\delta-\omega_{42})$ are the detunings, and $\delta$ is the
two-photon detuning.}%
\label{fig2}%
\end{figure}

\section{Frequency degenerate squeezed light}

In our experiment \cite{Jia}, the state $|g,m\rangle$ (or state $|1,2\rangle$)
involves the hyperfine levels $|5S_{1/2},F=2,3\rangle$, where the hyperfine
splitting of the ground state is $\omega_{21}=2\pi\times3.035$ GHz, and the
excited state $|e\rangle$ (or state $|3,4\rangle$) is $|5P_{1/2}\rangle$ has
an excited state decay rate of $\gamma=2\pi\times5.75$ MHz. The pump field is
blue-detuned approximately $1$ GHz to the D1 line of Rb-85 $5S_{1/2}%
\rightarrow5P_{1/2}$. The powers of the pump fields $E_{P1}$ and $E_{P2}$ are
set to $350$ mW, and their waists at the crossing point are 622 $\mu m$ and
596 $\mu m$, respectively. The Rabi frequencies of $\Omega_{P1}$ and
$\Omega_{P2}$ are $\Omega_{P1}\simeq28\gamma$ and $\Omega_{P2}\simeq30\gamma$
for a effective electric dipole $d=1.47\times10^{-29}$ Cm \cite{Steck}. The
atomic number density of Rb-85 at 125 $^{\circ}\mathrm{C}$ is approximately
$N\simeq4.5\times10^{18}$ m$^{-3}$. As shown in Fig.~\ref{fig1}(d), the pump
fields $P_{1}$ and $P_{2}$ have a certain angle $\theta_{0}$\ in one plane,
where the small angle $\theta_{0}$ can vary within a certain range. The probe
field $p$ is input at angles $\theta_{1}$\ and $\theta_{2}$\ relative to the
pump fields $P_{1}$ and $P_{2}$, respectively. The angle $\theta_{0}$
determines the minimum of the sum of angles $\theta_{1}$\ and $\theta_{2}$.
When only one pump field $P_{1}$ or $P_{2}$ exists, the conjugate field
$c_{1}$ or $c_{2}$ is generated under the respective phase matching condition
$2\mathbf{k}_{P_{1}}-n_{p}\mathbf{k}_{p}-\mathbf{k}_{c1}=0$ or $2\mathbf{k}%
_{P_{2}}-n_{p}\mathbf{k}_{p}-\mathbf{k}_{c2}=0$ \cite{McCormick07}. When two
pump fields $P_{1}$ and $P_{2}$ exist at the same time and changing the angles
$\theta_{1}$\ and $\theta_{2}$, under a certain condition a new conjugate
field $c$ is generated due to the new phase matching condition $\mathbf{k}%
_{P_{1}}+\mathbf{k}_{P_{2}}-n_{p}\mathbf{k}_{p}-\mathbf{k}_{c}=0$ \cite{Jia},
and the conjugate fields $c_{1}$ and $c_{2}$ disappear due to mismatching. The
probe field $p$ and the conjugate field $c$ are at an angle of $\theta_{3}$
relative to the $z$-axis, as shown in Fig.~\ref{fig1}(d).

Now, we use this configuration to produce a frequency degenerate two-mode
squeezed light field. The approach to generating the frequency degenerate twin
beams is based on the idea of inverting the configuration \cite{Corzo11}. Two
realtively strong beams\ is pumped the atomic system with the frequency of the
probe and conjugate beams, and along the direction of them, and a week beam
having the frequency and direction of the previous pump is also input.
However, this directly exchange of role of the pump and probe and conjugate
beams does not lead to the desired intensity difference squeezing and instead
we have found excess beams and noise due to the unexpected nonlinear process
as shown in Fig.~\ref{fig2}(a). We found it necessary to tune the detuning of
$\Delta_{1}$ and $\Delta_{2}$ as approximately $1$ GHz and $-2$ GHz to
suppress these extra processes, as shown in Fig.~\ref{fig2}(b). This detuned
choice is beneficial to the acquisition of frequency degenerate two-mode
squeezed light. The reason is that the noise on the two sides of the atomic
line is asymmetrical as Davis \textit{et al.} \cite{Davis} pointed out. On the
other hand, we reduce the gain by adjusting the temperature from 125 $^{\circ
}\mathrm{C}$ to 105 $^{\circ}\mathrm{C}$, so that the unexpected nonlinear
process can be suppressed. Using this configuration we have investigated the
generation of the frequency degenerate and spatial nondegenerate twin beams.

\begin{figure}[ptbh]
\includegraphics[scale=0.3,angle=0]{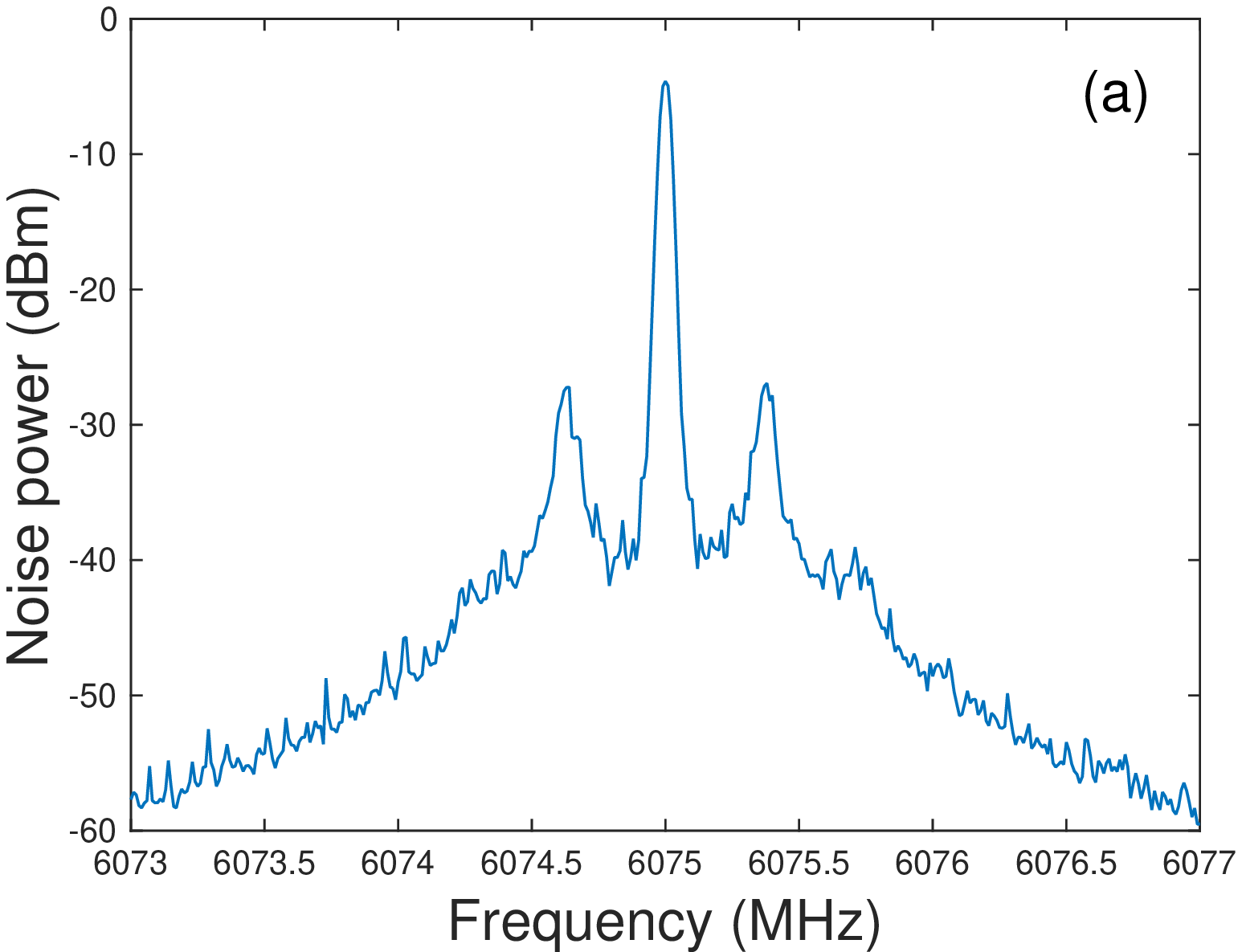}
\includegraphics[scale=0.3,angle=0]{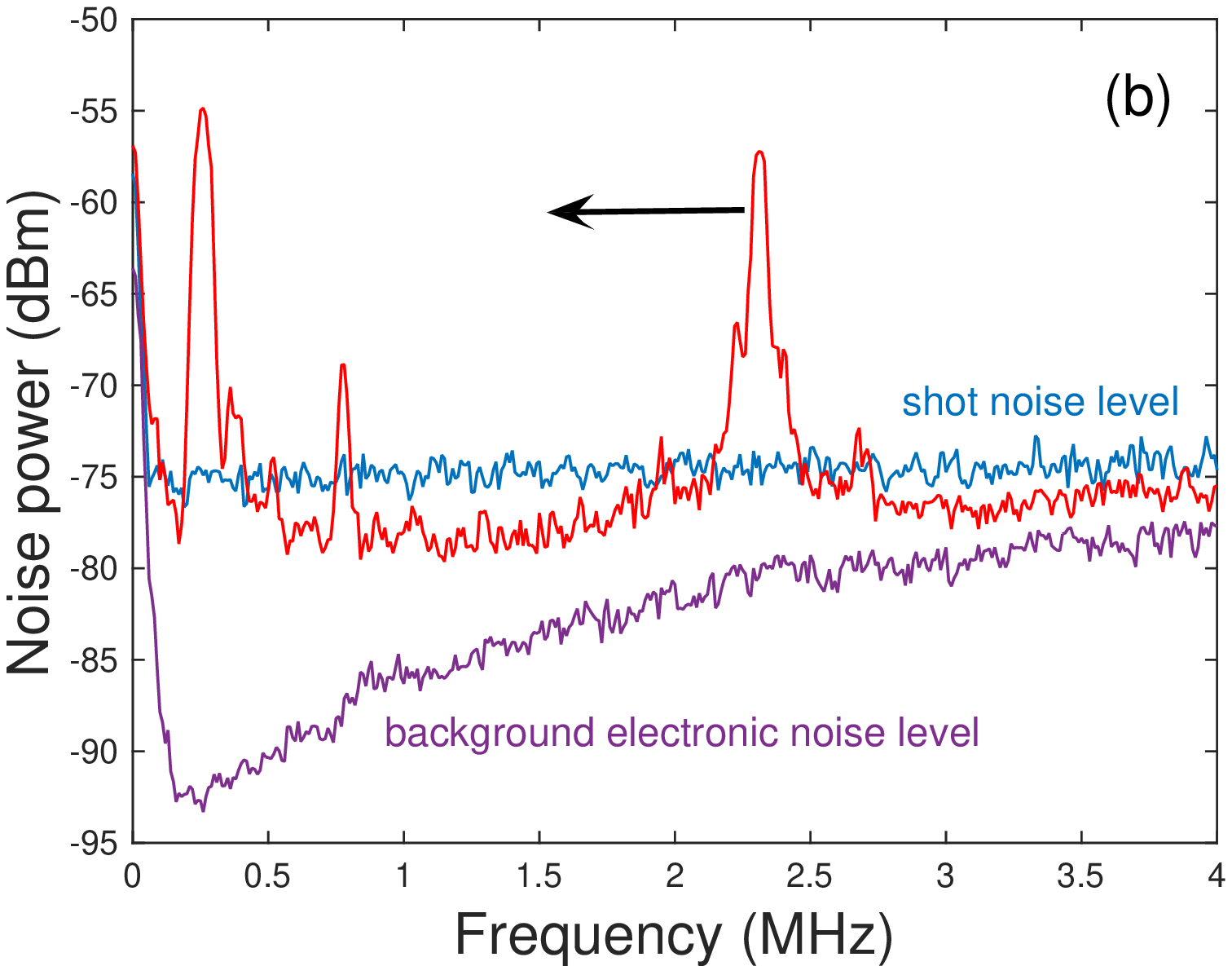}
\includegraphics[scale=0.3,angle=0]{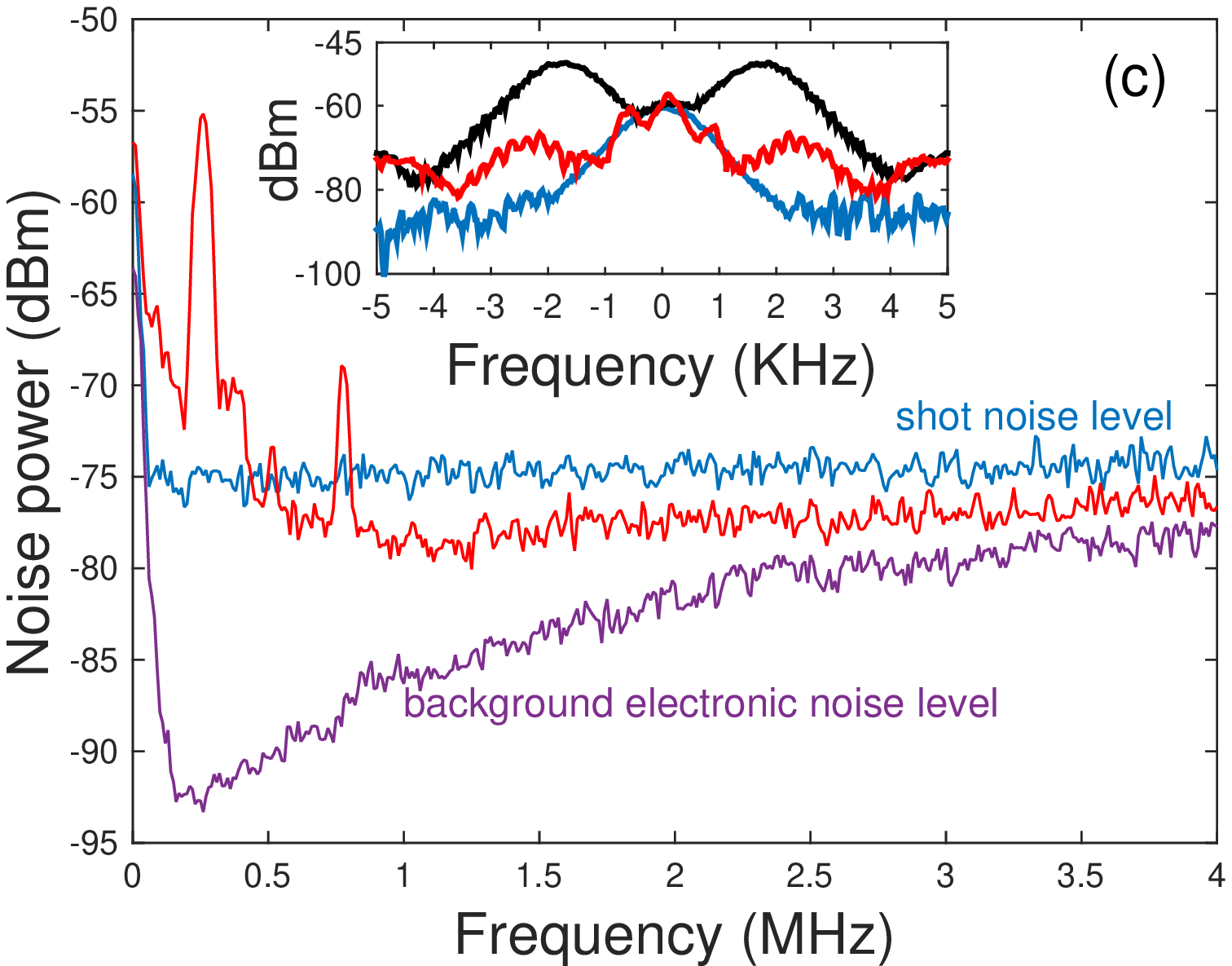} \caption{(a) Spectral noise
density of the beat signal produced by interference of two phase locked lasers
(pump1 and pump2). (b) Intensity-difference noise with frequency difference
between probe and conjugate beams $\sim2.3$ MHz. (c) Intensity-difference
noise with frequency difference between probe and conjugate beams $<1$ Hz.}%
\label{fig3}%
\end{figure}

We use the experimental setup with the two pump fields that are generated by a
Ti:Sapphire laser ($\Delta_{1}$ $\sim1$ GHz) and a semiconductor laser
($\Delta_{2}$ $\sim-2$ GHz) to implement this scheme, and the frequency
difference of the two pumps is achieved by using an OPLL with the beat
frequency of $6.075$ GHz as shown in Fig.~\ref{fig3}(a). The probe beam is
generated by frequency-shift the light form Ti: Sapphire with double-passed
$1.52$ GHZ acousto-optics modulators (AOM). The AOM frequency shift, and hence
the two-photon detuning $\delta$ is adjusted to optimize performance and
change the scheme from frequency non-degenerate to degenerate twin beams. In
the experiment of degenerate four-wave mixing, two pump lights with a
frequency difference of $6$ GHz or more are required to drive at the same
time. We use the amplifier lock scheme to generate pump light by frequency
shifting. In order to ensure a fixed phase difference between the two pumping
light fields, we use a beat frequency interlocking method-OPLL to lock a
semiconductor laser and a Ti:Sapphire laser to each other. Since the frequency
shift is as small as $0.1$ Hz, the lock relative frequency difference is less
than $1$ Hz.

The $6$ GHz beat frequency signal after the frequency locking system is
stabilized is shown in Fig.~\ref{fig3}(a). The modulating signal peaks
appearing in the frequency range of $\pm1$ MHz around both sides of the peak,
which is caused by the feedback noise of the OPLL system itself. When the
probe and conjugate beams are near degenerate and frequency difference between
them $\sim2.3$ MHz, the intensity-difference noise is shown in Fig.~\ref{fig3}%
(b), and we find the OPLL feedback noise become the main limitation to get
better squeezing. When the beat signal is further reduced, as shown by the
arrow in Fig.~\ref{fig3}(b), we obtain the intensity-difference noise with
frequency difference between probe and conjugate beams $<1$ Hz, as shown in
Fig.~\ref{fig3}(c), which indicate the twin beams are totally
indistinguishable . The inset in Fig.~\ref{fig3}(c) shows the process of
gradually reducing the beat signal to below $1$ Hz, where the black and red
lines are phase locked within less than $2$ KHz and $1$ Hz, respectively.

\section{Theoretical Model}

\begin{figure}[ptbh]
\centerline{\includegraphics[scale=0.8,angle=0]{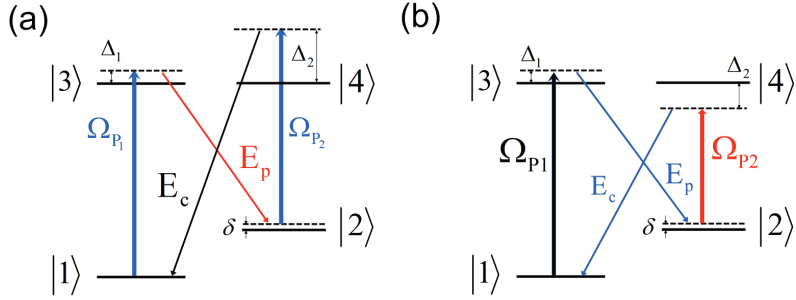}}\caption{(a)
Double-$\Lambda$ scheme with two pump fields $P_{1}$ and $P_{2}$. States
$|3\rangle$ and $|4\rangle$ are orthogonal linear combinations of magnetic
states of the excited hyperfine levels. $\Omega_{P1}$ and $\Omega_{P2}$ are
the Rabi frequencies, $\Delta_{1}$ $(=\omega_{P1}-\omega_{31})$ and
$\Delta_{2}$ $(=\omega_{P2}+\delta-\omega_{42})$ are the detunings, and
$\delta$ is the two-photon detuning. }%
\label{fig4}%
\end{figure}

In this section, we firstly describe the frequency non-degenerate squeezed
light based on non-collinear 4WM. As shown in Fig.~\ref{fig4}(a), we assume
that the two pump fields $E_{P1}$ and $E_{P2}$\ couple the transitions
$|1\rangle\rightarrow|3\rangle$ and $|2\rangle\rightarrow|4\rangle$,
respectively. The probe field couples the transition $|2\rangle\rightarrow
|3\rangle$, and the conjugate field couples the transition $|1\rangle
\rightarrow|4\rangle$. The transitions $|1\rangle\rightarrow|2\rangle$ and
$|3\rangle\rightarrow|4\rangle$ are not dipole allowed transitions. Since the
two pump fields $E_{P1}$ and $E_{P2}$ have the same polarizations and
frequencies, the two pump fields $E_{P1}$ and $E_{P2}$\ also couple the
transitions $|2\rangle\rightarrow|4\rangle$ and $|1\rangle\rightarrow
|3\rangle$ with probabilities of $1/2$. For the second set of coupling
transitions, we just swap the pump field $E_{P1}$ and the pump field $E_{P2}$
in the result for the first set of coupling transitions. We add the two sets
of conclusions to obtain the final result.

Next, we describe the frequency-degenerate squeezed light based on exchanging
the roles of the pump beams and the probe and conjugate beams. As shown in
Fig.~\ref{fig4}(b), the double-$\Lambda$\textbf{ }four-level process is the
same as the non-degenerate process except that the magnitude of the detuning
is different. Therefore, the frequency-degenerate and non-degenerate squeezed
lights based on non-collinear 4WM can be described by a same set of equations.

In the dipole and rotating wave approximations, the Hamiltonian of the atoms
combined with the Hamiltonian of the light-atom interaction is given by%
\begin{equation}
\hat{H}=\hat{H}_{atoms}+\hat{H}_{I},
\end{equation}
where%
\begin{equation}
\hat{H}_{atoms}=\hbar\omega_{41}\sigma_{44}+\hbar\omega_{31}\sigma_{33}%
+\hbar\omega_{21}\sigma_{22},
\end{equation}
and%
\begin{align}
&  \hat{H}_{I} =-\hbar(\Omega_{P1}e^{i(\mathbf{k}_{P1}\cdot\mathbf{r}%
-\omega_{P1}t)}\sigma_{31}+\Omega_{P2}e^{i(\mathbf{k}_{P2}\cdot\mathbf{r}%
-\omega_{P2}t)}\sigma_{42}\nonumber\\
&  +g_{p}\hat{\mathcal{E}}_{p}e^{i(\mathbf{k}_{p}\cdot\mathbf{r}-\omega_{p}%
t)}\sigma_{32}+g_{c}\hat{\mathcal{E}}_{c}e^{i(\mathbf{k}_{c}\cdot
\mathbf{r}-\omega_{c}t)}\sigma_{41})+\mathrm{H.c.}%
\end{align}
Here, $\omega_{n1}=\omega_{n}-\omega_{1}$ $(n=2,3,4)$, $\sigma_{nm}%
=|n\rangle\langle m|$ $(n,m=1,2,3,4)$, $2\Omega_{P1}=\mu_{31}\mathcal{E}%
_{P1}/\hbar$ and $2\Omega_{P2}=\mu_{42}\mathcal{E}_{P2}/\hbar$ are the Rabi
frequencies, $g_{n}$ ($n=p,c$)\ are the atom-field coupling constants, and
$\hat{\mathcal{E}}_{p}$ and $\hat{\mathcal{E}_{c}}$\ are the slowly varying
envelope operators of the probe and conjugate field.

\begin{figure}[ptbh]
\centerline{\includegraphics[scale=0.4,angle=0]{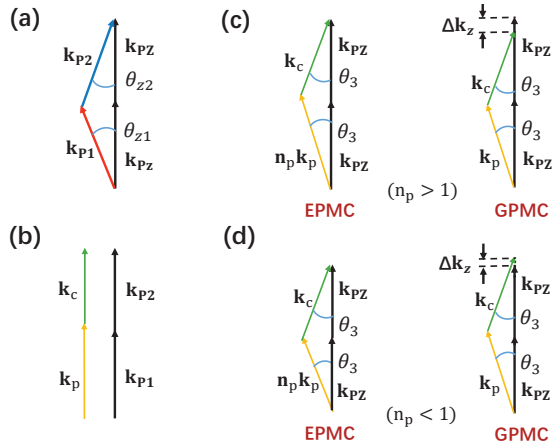}} \caption{(a) The
wavevectors $\mathbf{k}_{P1}$ and $\mathbf{k}_{P2}$ are projected onto the
axis as $2\mathbf{k}_{PZ}$. The angles between the $\mathbf{k}_{PZ}$ and the
pump fields $P_{1}$ and $P_{2}$ are $\theta_{z1}$\ and $\theta_{z2}$,
respectively. (b) The configuration where the geometric phase matching
condition (GPMC) is fulfilled: $\Delta k_{z}=0$. The configuration where the
effective phase matching condition (EPMC) is fulfilled ($2\left\vert
\mathbf{k}_{PZ}\right\vert -n_{p}\left\vert \mathbf{k}_{p}\right\vert
\cos\theta_{3}-\left\vert \mathbf{k}_{c}\right\vert \cos\theta_{3}=0$) for an
effective index of refraction of the probe (c) $n_{p}>1$ and (d) $n_{p}<1$,
with necessary geometric mismatches of (c) $\Delta k_{z}>0$ and (d) $\Delta
k_{z}<0$ with $\Delta k_{z}=2\left\vert \mathbf{k}_{PZ}\right\vert -\left\vert
\mathbf{k}_{p}\right\vert \cos\theta_{3}-\left\vert \mathbf{k}_{c}\right\vert
\cos\theta_{3}$, respectively.}%
\label{fig5}%
\end{figure}

The equations for the atomic operators $\sigma_{nm}$ ($n,m=1,2,3,4)$ in the
Heisenberg picture are given in the Appendix. Using the atomic operators to
evaluate the linear and nonlinear components of the polarization at
$\omega_{p}$ and $\omega_{c}$, the polarization of the atomic medium at a
particular frequency is given by $\hat{P}\left(  \omega_{p}\right)
=Nd_{23}\tilde{\sigma}_{23}+\mathrm{H.c.}$ and $\hat{P}\left(  \omega
_{c}\right)  =Nd_{14}\tilde{\sigma}_{14}+\mathrm{H.c.}$, where $N$\ is the
number density of the atomic medium. The polarizations of the medium at
frequency $\omega_{n}$ ($n=p,c$) are given by
\begin{align}
\hat{P}_{p}(\omega_{p})  &  =\sqrt{\frac{\hbar\omega_{p}}{2\epsilon_{0}V}%
}\epsilon_{0}\chi_{pp}(\omega_{p})\hat{\mathcal{E}}_{p}e^{i\mathbf{k}_{p}%
\cdot\mathbf{r}}+\sqrt{\frac{\hbar\omega_{c}}{2\epsilon_{0}V}}\epsilon_{0}%
\chi_{pc}(\omega_{p})\hat{\mathcal{E}}_{c}^{\dag}e^{i(\mathbf{k}%
_{P1}+\mathbf{k}_{P2}-\mathbf{k}_{c})\cdot\mathbf{r}}+\mathrm{H.c.},\\
\hat{P}_{c}(\omega_{c})  &  =\sqrt{\frac{\hbar\omega_{c}}{2\epsilon_{0}V}%
}\epsilon_{0}\chi_{cc}(\omega_{c})\hat{\mathcal{E}}_{c}e^{i\mathbf{k}_{c}%
\cdot\mathbf{r}}+\sqrt{\frac{\hbar\omega_{p}}{2\epsilon_{0}V}}\epsilon_{0}%
\chi_{cp}(\omega_{c})\hat{\mathcal{E}}_{p}^{\dag}e^{i(\mathbf{k}%
_{P1}+\mathbf{k}_{P2}-\mathbf{k}_{p})\cdot\mathbf{r}}+\mathrm{H.c.}.
\end{align}
Here, the two coefficients $\chi_{pp}$ and $\chi_{cc}$ describe the effective
linear polarization processes for the probe and conjugate fields,
respectively, and unlike the usual linear coefficients, they depend
nonlinearly on the pump field. The other two coefficients $\chi_{pc}$ and
$\chi_{cp}$ are responsible for the 4WM process. A detailed calculation is
given in the Appendix.

\begin{figure}[ptbh]
\includegraphics[scale=0.35,angle=0]{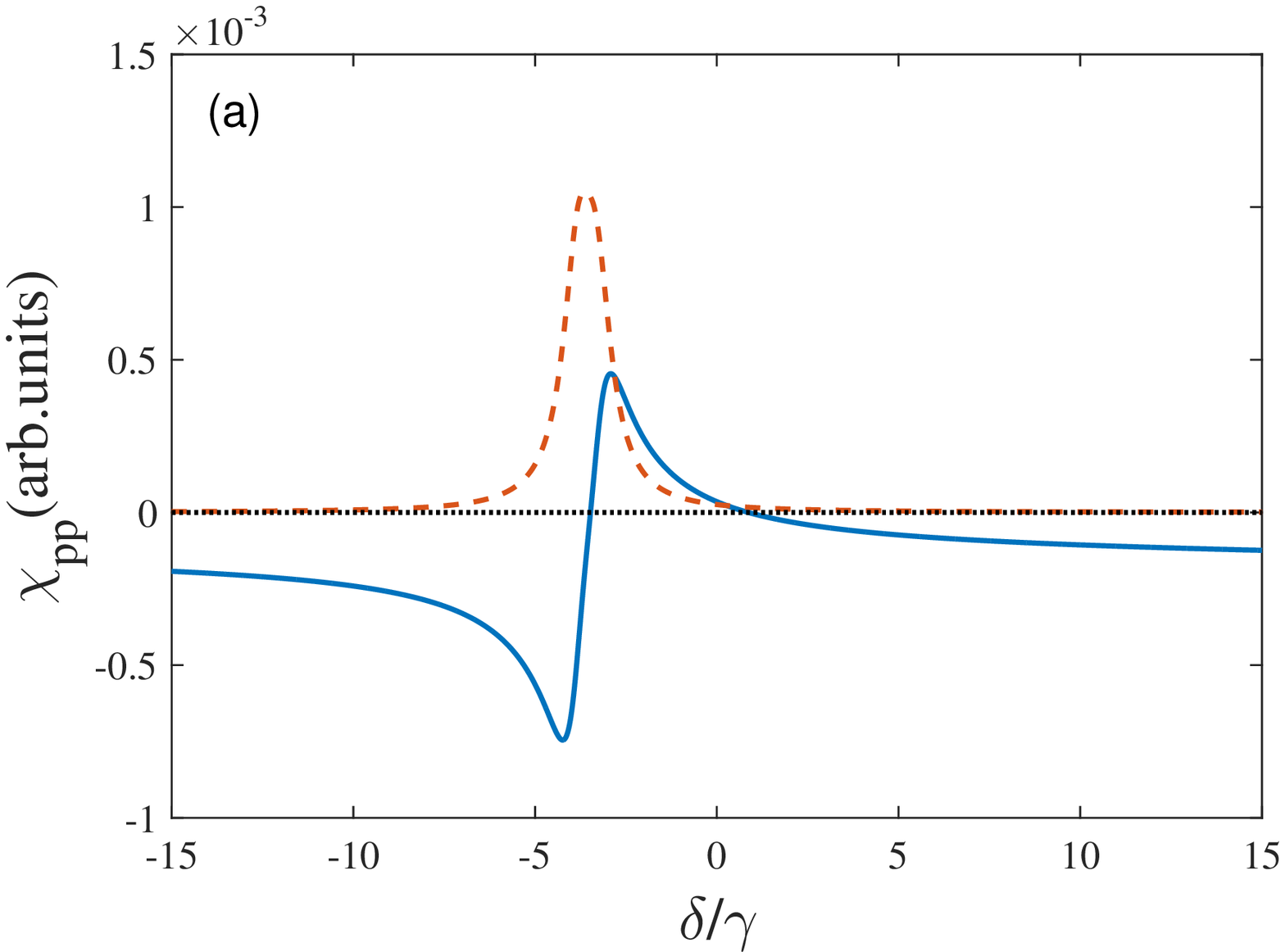}
\includegraphics[scale=0.35,angle=0]{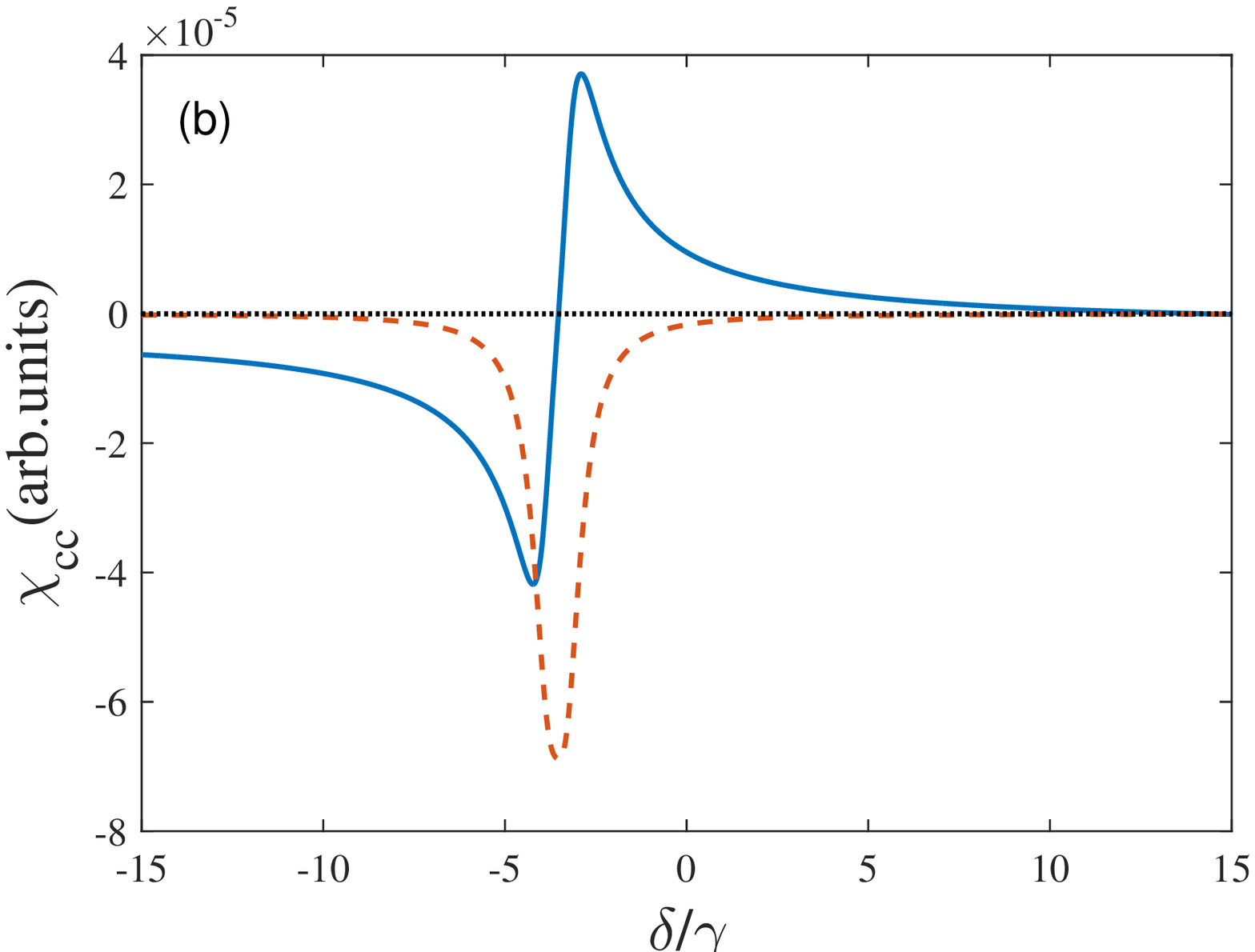}
\caption{The direct and cross susceptibilities for the probe and conjugate
fields as a function of the two-photon detuning $\delta/\gamma$. The solid
lines are the real parts, and the dashed lines are the imaginary parts. The
excited state decay rate is $\gamma=2\pi\times5.75$ MHz, and $\gamma
_{c}=0.5\gamma$. The hyperfine splitting of the ground state is $\omega
_{21}=2\pi\times3.035$ GHz. The detuning of pump1 is $\Delta_{1}=174\gamma$.
The Rabi frequencies of $\Omega_{P1}$ and $\Omega_{P2}$ are $\Omega
_{P1}=28\gamma$ and $\Omega_{P2}=30\gamma$, respectively.}%
\label{fig6}%
\end{figure}

Under the condition of the slowly varying amplitude approximation, considering
nearly co-propagating beams along the $z$ axis, these field equations in the
co-moving frame are written as%
\begin{align}
\frac{\partial}{\partial z}\hat{\mathcal{E}}_{p}  &  =\frac{ik_{p}}{2}%
[\chi_{pp}(\omega_{p})\hat{\mathcal{E}}_{p}+\chi_{pc}^{\prime}(\omega_{p}%
)\hat{\mathcal{E}}_{c}^{\dag}e^{i\Delta k_{z}z}],\label{pro1}\\
\frac{\partial}{\partial z}\hat{\mathcal{E}}_{c}  &  =\frac{ik_{c}}{2}%
[\chi_{cc}(\omega_{c})\hat{\mathcal{E}}_{c}+\chi_{cp}^{\prime}(\omega_{c}%
)\hat{\mathcal{E}}_{p}^{\dag}e^{i\Delta k_{z}z}], \label{pro2}%
\end{align}
where $\chi_{pc}^{\prime}=\chi_{pc}\sqrt{\omega_{c}/\omega_{p}}\simeq\chi
_{pc}$, $\chi_{cp}^{\prime}=\chi_{cp}\sqrt{\omega_{p}/\omega_{c}}\simeq
\chi_{cp}$, and $\Delta k_{z}$ is the projection of the geometric phase
mismatch $\Delta\mathbf{k}=\mathbf{k}_{P_{1}}+\mathbf{k}_{P_{2}}%
-\mathbf{k}_{p}-\mathbf{k}_{c}$ on the $z$\ axis. The solutions to the
propagation equations (\ref{pro1})\ and (\ref{pro2}) with a medium of length
$L$ are given by%
\begin{align}
\hat{\mathcal{E}}_{p}  &  =G_{1}\hat{\mathcal{E}}_{p}(0)+g_{1}\hat
{\mathcal{E}}_{c}^{\dag}(0),\label{solu1}\\
\hat{\mathcal{E}}_{c}^{\dag}  &  =\left[  G_{2}\hat{\mathcal{E}}_{c}^{\dag
}(0)+g_{2}\hat{\mathcal{E}}_{p}(0)\right]  e^{-i\Delta k_{z}L}, \label{solu2}%
\end{align}
where%
\begin{align}
G_{1}  &  =e^{\delta aL}[\cosh(\xi L)+\frac{a}{\xi}\sinh(\xi L)],\text{ }%
G_{2}=e^{\delta aL}[\cosh(\xi L)-\frac{a}{\xi}\sinh(\xi L)],\nonumber\\
g_{1}  &  =\frac{a_{pc}}{\xi}e^{\delta aL}\sinh(\xi L),\text{ }g_{2}%
=-\frac{a_{cp}}{\xi}e^{\delta aL}\sinh(\xi L), \label{gain}%
\end{align}
and%
\begin{align}
a_{pj}  &  =ik_{p}\chi_{pj}/2,\text{ }a_{cj}=ik_{c}\chi_{cj}^{\ast}/2\text{
}(j=p,c),\nonumber\\
a  &  =(a_{pp}+a_{cc}-i\Delta k_{z})/2,\text{ }\xi=\sqrt{a^{2}-a_{pc}a_{cp}%
},\text{ }\delta a=(a_{pp}-a_{cc}+i\Delta k_{z})/2.
\end{align}

\begin{figure}[ptbh]
\centerline{\includegraphics[scale=0.41,angle=0]{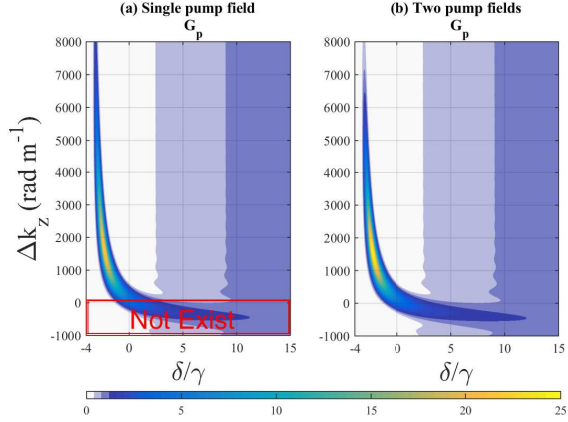}}\caption{Theoretical
output probe gain $G_{p}$ as a function of the two-photon detuning
$\delta/\gamma$ and the geometrical phase match $\Delta k_{z}$ with (a) a
single pump field and (b) two pump fields. Here, the excited state decay rate
is $\gamma=2\pi\times5.75$ MHz, the decoherence rate is $\gamma_{c}=0.5\gamma
$, the atom density is $N=4.5\times10^{18}$ m$^{-3}$, the length of the medium
is $L=12.5$ mm and the pump Rabi frequencies are $\Omega_{P1}=28\gamma$ and
$\Omega_{P2}=30\gamma$.}%
\label{fig75}%
\end{figure}

The number operators of the probe beam and conjugate beam are defined as
$\hat{N}_{p}=\hat{\mathcal{E}}_{p}^{\dag}\hat{\mathcal{E}}_{p}$ and $\hat
{N}_{c}=\hat{\mathcal{E}}_{c}^{\dag}\hat{\mathcal{E}}_{c}$, respectively. From
the above result, we define the gain of the probe beam in the 4WM process as:
\begin{equation}
G_{p}=\frac{\langle\hat{N}_{p}\rangle_{out}}{\langle\hat{N}_{p}\rangle_{in}%
}\simeq\left\vert G_{1}\right\vert ^{2}, \label{Gp}%
\end{equation}
where the initial condition is $\langle\hat{N}_{p}\rangle_{in}\gg1$ and
$\langle\hat{N}_{c}\rangle_{in}=0$. The 4WM generates a correlated probe and
conjugate beams, and the relative intensity fluctuations are reduced for the
amplification process. After the 4WM, the relative intensity fluctuation is
given by%
\begin{equation}
\Delta^{2}(\hat{N}_{p}-\hat{N}_{c})_{out}=(\left\vert G_{1}\right\vert
^{2}-\left\vert g_{2}\right\vert ^{2})^{2}\Delta^{2}(\hat{N}_{p}%
)_{in}+\left\vert g_{1}^{\ast}G_{1}-g_{2}^{\ast}G_{2}\right\vert ^{2}%
[\langle\hat{N}_{p}\rangle_{in}+1].
\end{equation}
Hence the beams have been amplified without increasing the relative intensity
noise, and they are relative intensity squeezed. The standard quantum limit
(SQL) is a differential measurement equal to the total optical power, that is%
\begin{equation}
\langle\hat{N}_{p}-\hat{N}_{c}\rangle_{SQL}\equiv\langle\hat{N}_{p}+\hat
{N}_{c}\rangle\simeq(\left\vert G_{1}\right\vert ^{2}+\left\vert
g_{2}\right\vert ^{2})\langle\hat{N}_{p}\rangle_{in}.
\end{equation}
The noise figure of the process (or \textquotedblleft degree of
squeezing\textquotedblright) is the ratio of the measured noise to the
corresponding shot-noise level for equal optical power. The typically the
noise figure is quoted as the noise in decibels relative to the SQL.

\section{Phase matching}

\begin{figure}[ptbh]
\centerline{\includegraphics[scale=0.45,angle=0]{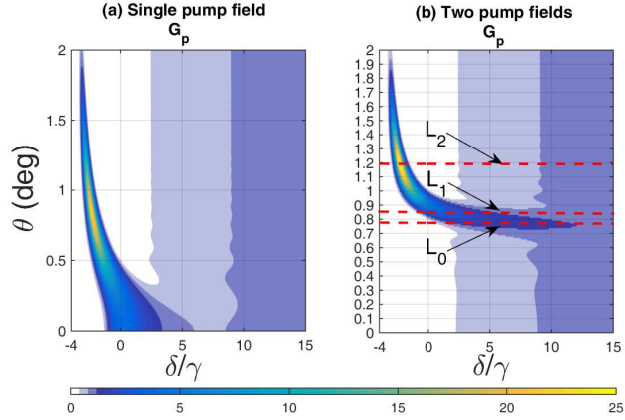}}
\caption{Theoretical output probe gain $G_{p}$ as a function of the two-photon
detuning $\delta/\gamma$ and the probe-pump angle $\theta$ with (a) a single
pump field and (b) two pump fields. The area intersecting the dashed line is
the area selected by our experimental parameters. The angle between pump
fields $P_{1}$ and $P_{2}$ is $\theta_{0}/2=0.615^{\mathrm{o}}$. The angles
between the probe field and the pump fields $P_{1}$ and $P_{2}$ are
$\theta_{1}=\theta_{2}=\theta=0.861^{\mathrm{o}}$.}%
\label{fig8}%
\end{figure}

In this section, we describe the angles $\theta_{1}$\ and $\theta_{2}$ between
the probe field and the pump fields by phase matching based on the different
refractive indices.

As shown in Fig. \ref{fig5}(a), when two pump fields $E_{P1}$ and $E_{P2}$ are
incident at an angle, the total projection of the wavevector of the pump
fields onto the $z$-axis is $2\mathbf{k}_{PZ}$ and becomes smaller; i.e.,
$2\left\vert \mathbf{k}_{PZ}\right\vert <\left\vert \mathbf{k}_{P1}\right\vert
+\left\vert \mathbf{k}_{P2}\right\vert $. The geometric phase matching
condition (GPMC) is given by%
\begin{align}
\Delta k_{z}  &  =\left\vert \mathbf{k}_{P1}\right\vert \cos(\theta
_{z1})+\left\vert \mathbf{k}_{P2}\right\vert \cos(\theta_{z2})-(\left\vert
\mathbf{k}_{p}\right\vert +\left\vert \mathbf{k}_{c}\right\vert )\cos
\theta_{3}\nonumber\\
&  =2\left\vert \mathbf{k}_{PZ}\right\vert -(\left\vert \mathbf{k}%
_{p}\right\vert +\left\vert \mathbf{k}_{c}\right\vert )\cos\theta_{3},
\label{GPM}%
\end{align}
where $\theta_{3}$ is the angle between the probe and the projected pump
field. In fact, if the 4WM is efficient, the GPMC of Eq. (\ref{GPM}) may not
be satisfied, but the effective phase matching condition (EPMC) must be met:%
\begin{equation}
\mathbf{k}_{P1}+\mathbf{k}_{P2}-n_{p}\mathbf{k}_{p}-n_{c}\mathbf{k}_{c}=0,
\label{EPM}%
\end{equation}
where the refractive index $n_{p}=\sqrt{1+\operatorname{Re}(\chi_{pp})}$, and
$n_{c}=\sqrt{1+\operatorname{Re}(\chi_{cc})}$. For the case of two pump
fields, the Eq. (\ref{EPM}) can be written as%
\begin{equation}
\cos\theta_{3}=\frac{\omega_{P1}\cos(\theta_{z1})+\omega_{P2}\cos(\theta
_{z2})}{n_{p}\omega_{p}+n_{c}\omega_{c}}. \label{theta3}%
\end{equation}
According to $\theta_{3}$ and $\theta_{z1}$ ($\theta_{z2}$), we can determine
the angle $\theta_{1}$ ($\theta_{2}$) between the probe field and the pump
field $P_{1}$ ($P_{2}$).

In order to better explain phase matching, we first consider the
non-degenerate 4WM case. For the case of two degenerate pump fields, the
conservation of energy impose the condition $\omega_{p}+\omega_{c}=\omega
_{P1}+\omega_{P2}=2\omega_{0}$, where $\omega_{0}$ is the frequency of the
pump field. Considering $\theta_{z1}=\theta_{z2}=\theta_{0}/2$, the Eq.
(\ref{theta3}) is written as
\begin{equation}
\cos\theta_{3}=\frac{2\omega_{0}\cos(\theta_{0}/2)}{n_{p}\omega_{p}+\omega
_{c}}, \label{EPM2}%
\end{equation}
where $n_{c}\simeq1$ due to the conjugate field with a large detuning. For a
given angle $\theta_{0}$, when $n_{p}=1$, the EPMC of Eq. (\ref{EPM2}) imposes
$\theta_{3}=\theta_{0}/2$. Under this condition, the GPMC $\Delta k_{z}=0$ is
also satisfied, which is the phase matching condition in free space, where the
beams are required rigorously copropagating as shown in Fig. \ref{fig5}(b).

When $n_{p}>1$, the EPMC of Eq. (\ref{EPM2}) is established to require that
$\theta_{3}>\theta_{0}/2$, which means that the GPMC of Eq. (\ref{GPM}) cannot
be satisfied and will occur $\Delta k_{z}>0$, as shown in Fig. \ref{fig5}(c).
Considering $\theta_{1}=\theta_{2}=\theta$, using the law of cosines we obtain
the angle requirement between the probe field and the pump fields:
\begin{equation}
\theta>\cos^{-1}\left[  \frac{1+\cos\theta_{0}}{2}\right]  .
\end{equation}

If $n_{p}<1$, similarly, the EPMC of Eq. (\ref{EPM2}) requires that
$\theta_{3}<\theta_{0}/2$ and imposes $\Delta k_{z}<0$, as shown in Fig.
\ref{fig5}(d). In addition, the generated probe and conjugate beams have
separate directions, which requires that the angle $\theta_{3}>0$.
Furthermore, using the minimum value of the refractive index $\min(n_{p}%
)$\ according to Eq. (\ref{EPM2}), we obtain%
\begin{equation}
0<(\theta_{3})_{\min}<\theta_{3}<\frac{\theta_{0}}{2},
\end{equation}
where $(\theta_{3})_{\min}=\cos^{-1}\left[  2\omega_{0}\cos(\theta
_{0}/2)/(\min(n_{p})\omega_{p}+\omega_{c})\right]  $ and correspondingly,%
\begin{equation}
\frac{\theta_{0}}{2}<\theta_{\min}<\theta<\cos^{-1}\left[  \frac{1+\cos
\theta_{0}}{2}\right]  , \label{angle1}%
\end{equation}
where $\theta_{\min}=\cos^{-1}\left[  \omega_{0}(1+\cos\theta_{0})/(\min
(n_{p})\omega_{p}+\omega_{c})\right]  $. Compared to the single pump field
case, this is a new region. Because for the single pump field case, the angle
$\theta_{0}$ here is equivalent to $0$, where the condition $\theta_{3}%
<\theta_{0}/2$ cannot be satisfied because $\theta_{3}$ cannot be less than
$0$. That is, when $n_{p}<1$, the EPMC cannot be satisfied for 4WM driven by a
single pump field.

For degenerate case, the form of Eqs. (\ref{GPM}-\ref{theta3}) is the same
except for the magnitude of the wave vectors. With two strong beams with the
frequency of the probe and conjugate beams, and along the direction of them,
and a week beam having the frequency and direction of the previous pump, we
can generate of the frequency degenerate and spatial nondegenerate twin beams
by changing the detunings of $\Delta_{1}$ and $\Delta_{2}$. Similar to
non-degenerate case, if $n_{p}+n_{c}=2$, then GPMC $\Delta k_{z}=0$ is
satisfied. If $n_{p}+n_{c}>2$ or $n_{p}+n_{c}<2$, then the corresponding GPMC
$\Delta k_{z}>0$ or $\Delta k_{z}<0$ is also obtained.

\section{Data analysis}

In this section, we numerically analyze the characteristics of the squeezed
light produced in this new region with $n_{p}<1$.

Fig.~\ref{fig6}(a-b) show the direct susceptibilities $\chi_{pp}$ and
$\chi_{cc}$ for the probe and conjugate fields as a function of the two-photon
detuning $\delta/\gamma$, and we obtain that $\chi_{cc}$ is far less than
$\chi_{pp}$ due to large detuning. In Fig.~\ref{fig6}(a), when $\delta<0$, the
real part of $\chi_{pp}$ is effectively responsible for the index of
refraction of the probe for the single pump field case
\cite{Jasperse11,Turnbull13}. However, for our two pump fields input case, the
phase matching condition can also be satisfied on the other side $\delta>0$.

According to Eq. (\ref{Gp}), we plot the probe gain $G_{p}$ as a function of
the two-photon detuning $\delta$\ and the geometric phase mismatch $\Delta
k_{z}$ in the presence of a single pump field and two pump fields, as shown in
Fig. \ref{fig7}. One can see that the maximum gains are obtained on the side
$\delta<0$ with $\Delta k_{z}>0$, for both cases. When $\Delta k_{z}<0$, the
probe gain $G_{p}$ does not exist for the single pump field input case and
occurs for two pump fields input case. The bandwidth of the probe gain is
relatively large due to the slow change in the refractive index with the
two-photon detuning.

\begin{figure}[ptbh]
\includegraphics[scale=0.4,angle=0]{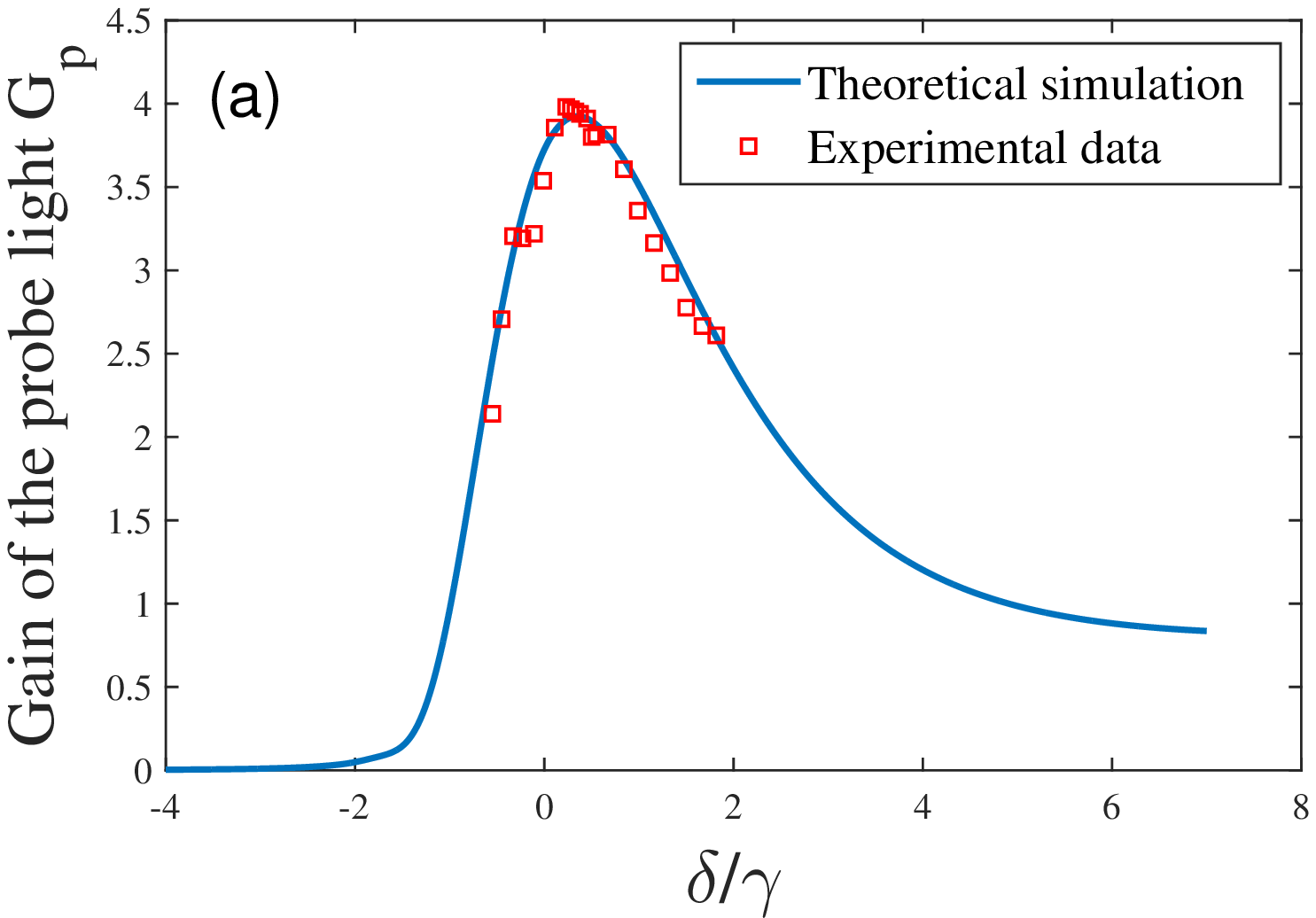}
\includegraphics[scale=0.4,angle=0]{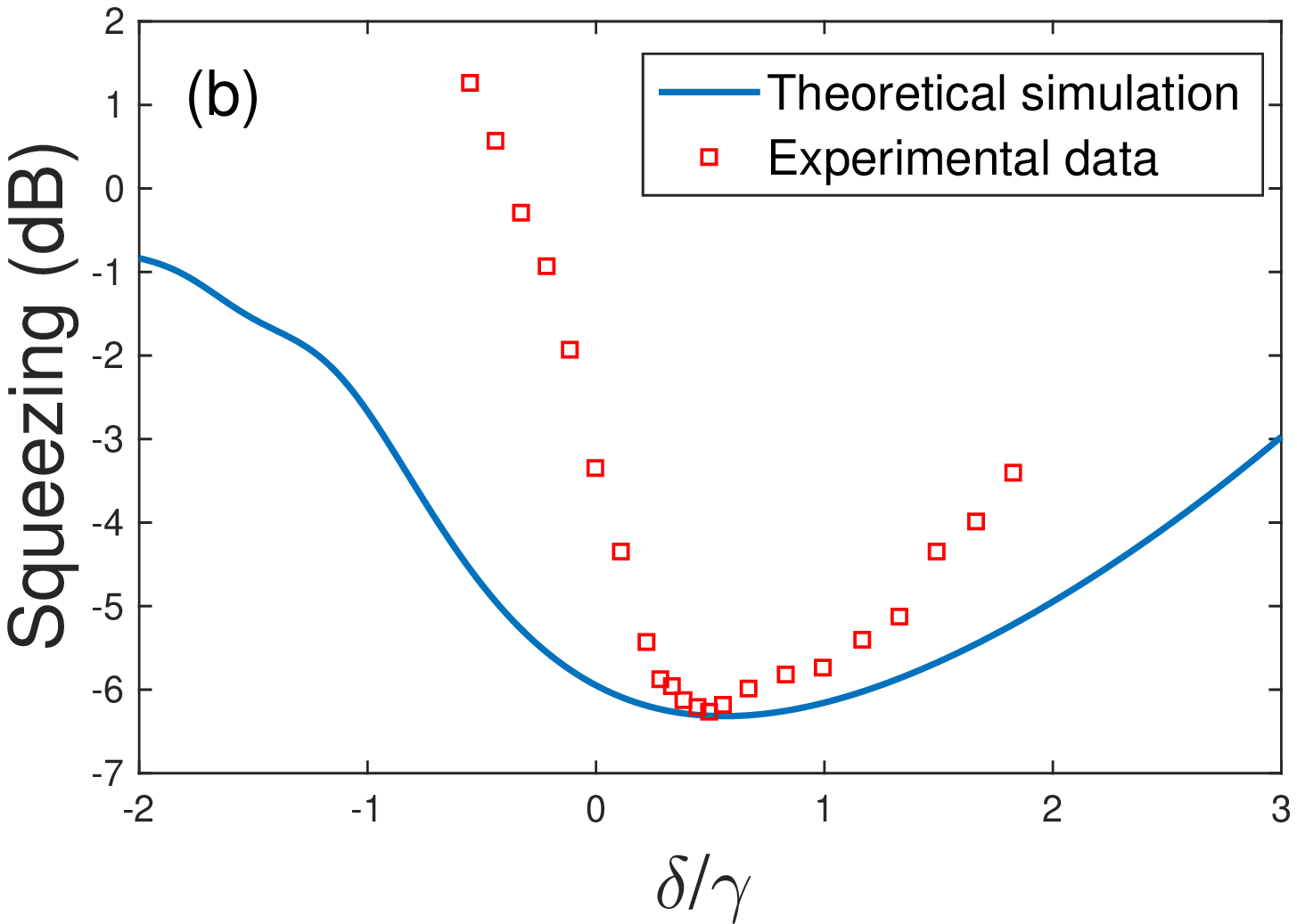} \caption{ (a) The gain of the
probe field and (b) the squeezing as a function of the two-photon detuning
$\delta$. The parameters are as follows: $\gamma=2\pi\times5.75$ MHz,
$\gamma_{c}=0.5\gamma$, $N=4.5\times10^{18}$ m$^{-3}$, $L=12.5$ mm,
$\Omega_{P1}=28\gamma$, $\Omega_{P2}=30\gamma$, $\theta_{0}%
/2=0.615^{\mathrm{o}}$, and $\theta_{1}=\theta_{2}=0.861^{\mathrm{o}}$.}%
\label{fig9}%
\end{figure}

The theoretical output probe gain $G_{p}$ as a function of the two-photon
detuning $\delta/\gamma$ and the probe-pump angle $\theta$ with (a) a single
pump field and (b) two pump fields is shown in Fig. \ref{fig8}, where we
consider $\mathbf{k}_{P1}=\mathbf{k}_{P2}=\mathbf{k}_{P}$ and $\theta
_{1}=\theta_{2}=\theta$. It can be seen from that for a single pump field, the
gain and 4WM process is on the $\delta<0$ side as the angle $\theta$ increases
due to phase matching. For the case of two pump fields, the gain and 4WM
process can be achieved on the left side (Line $L_{2\text{ }}$) or the right
side (line $L_{0\text{ }}$or line $L_{1\text{ }}$)\ by choosing the angle
$\theta$ between the probe field and the pump fields, for a given angle
$\theta_{0}$.

The area intersecting the dashed line $L_{1\text{ }}$in Fig. \ref{fig8}(b) is
the area selected by our experimental parameters, where $\theta_{0}%
/2=0.615^{\mathrm{o}}$, and $\theta=$ $0.861^{\mathrm{o}}$. According to the
minimum value $\min(\operatorname{Re}(\chi_{pp}))=-1.609\times10^{-4}$ in Fig.
\ref{fig6}, using Eq. (\ref{angle1}), we obtain
\begin{equation}
0.7630^{\mathrm{o}}<\theta<0.8697^{\mathrm{o}}.
\end{equation}

If we only choose the angle $\theta$ based on the bandwidth, we choose line
$L_{0\text{ }}$[Fig. \ref{fig8}(b)] because it has the largest bandwidth.
However, in the experiment, the angle is finely adjusted according to the
degree of squeezing, and the optimum value of the angle $\theta$ is different.
If the angle $\theta$ is chosen as $1.2^{\mathrm{o}}$ of line $L_{2\text{ }}%
$in Fig. \ref{fig8}(b), the 4WM process driven by two pump fields will also be
observed on the $\delta<0$ side due to the large gain. As shown in Fig.
\ref{fig7}, the strong conjugate field $c$ and two weak conjugate fields
$c_{1}$ and $c_{2}$ may all occur because of their gains \cite{Knutson18}.
However, in this region, the absorption is also large, which will affect the
degree of squeezing of the two generated beams.

Here, $\theta_{0}$ has a fundamental effect on wavevector matching in the new
4WM process, thus opening up a region in which high-intensity-difference
squeezed light can be obtained over a wide bandwidth with low loss and
moderate gain. The gain curve in Fig. \ref{fig8}(b) shifts upward as the angle
$\theta_{0}$ increases, because the minimum value of the angle $\theta$ is
greater than the angle $\theta_{0}/2$.

The gain of the probe field and the squeezing as a function of the two-photon
detuning $\delta$ is shown in Fig. \ref{fig9}, where the square represents
experimental data and the solid line is a theoretical simulation. The
theoretical simulations and experimental data of the gain of the probe field
are in good agreement as shown in Fig. \ref{fig9}(a). The squeezing degree is
affected by the spatial mode mismatch, optical absorption by atomic system,
optical loss in the light path, and atomic decoherence. Fig. \ref{fig9}(b)
shows the theoretical simulations and experimental data of squeezing, where
the theoretical squeezing curve is reduced by 0.56 times and the agreement is
not very good because these effects are not included in our model in order to
clarify the physics picture concisely. On the new $\delta>0$ side as shown in
Fig. \ref{fig9}(a), the bandwidth of the gain is relatively larger than that
for the single pump field case, which is advantageous for realizing
wide-bandwidth frequency-degenerate and nondegenerate intensity-squeezed
light. These light fields can be widely used in quantum information and other fields.

\section{Conclusion}

We have studied that a novel two-mode squeezed light is generated from a 4WM
process driven by two pump fields crossing a small angle, where the twin beams
are generated with a new phase matching condition. Different from 4WM realized
by a single pump field where the gain peak can only be achieved on the
$\delta<0$ side, the new 4WM process is implemented from the $\delta<0$ side
to the $\delta>0$ side by an angle adjustment. The refractive index of the
corresponding probe field $n_{p}$ can be converted from $n_{p}>1$ to $n_{p}%
<1$, which can also be used to convert between slow light \cite{Boyer07} and
fast light \cite{Glasser12}. Based on slow light and fast light of the probe
field, two different time-order output pulses can be achieved. On the new
$\delta>0$ side, the refractive index $n_{p}$ changes slowly with the
two-photon detuning $\delta$ over a large range, which leads to a relatively
large gain bandwidth. With two strong beams with the frequency of the probe
and conjugate beams, and along the direction of them, and a week beam having
the frequency and direction of the previous pump, we have generated the
frequency degenerate and spatial nondegenerate twin beams with tuning the
detuning of $\Delta_{1}$ and $\Delta_{2}$. This type of twim beams can be
combined and interfered directly on the beam splitter. These wide-bandwidth
intensity-squeezed light fields can be applied in quantum information and
quantum metrology.

\section*{Funding}

C.-H.Y. is supported by the National Natural Science Foundation of China
(NSFC) under Grant No. 11474095 and the Fundamental Research Funds for the
Central Universities. L.-Q.C. is supported by the NSFC under Grant Nos.
11874152, 11604069, and 91536114 and the National Science Foundation of
Shanghai (No. 17ZR1442800). W.Z. is supported by the National Key Research and
Development Program of China under Grant No. 2016YFA0302001 and NSFC Grants
Nos. 11654005 and 11234003.

\appendix
\setcounter{equation}{0}
\renewcommand{\theequation}{S\arabic{equation}}

\section{Susceptibilities with two pump fields crossing a small angle input}

As shown in Fig.~\ref{fig3}, we assume that the two pump fields $E_{P1}$ and
$E_{P2}$\ couple the transitions $|1\rangle\rightarrow|3\rangle$ and
$|2\rangle\rightarrow|4\rangle$, respectively. The probe field couples the
transition $|2\rangle\rightarrow|3\rangle$, and the conjugate field couples
the transition $|1\rangle\rightarrow|4\rangle$. The transitions $|1\rangle
\rightarrow|2\rangle$ and $|3\rangle\rightarrow|4\rangle$ are not dipole
allowed transitions.

Consequently, we obtain the following set of equations for the populations
$\sigma_{nn}$:%
\begin{align}
&  \frac{\partial\sigma_{11}}{\partial t}=\Gamma_{13}\sigma_{33}+\Gamma
_{14}\sigma_{44}+i(\Omega_{P1}^{\ast}e^{-i\mathbf{k}_{P1}\cdot\mathbf{r}%
}\tilde{\sigma}_{13}+g_{c}\hat{\mathcal{E}}_{c}^{\dagger}e^{-i\mathbf{k}%
_{c}\cdot\mathbf{r}}\tilde{\sigma}_{14}-\Omega_{P1}e^{i\mathbf{k}_{P1}%
\cdot\mathbf{r}}\tilde{\sigma}_{31}\nonumber\\
&  -g_{c}\hat{\mathcal{E}}_{c}e^{i\mathbf{k}_{c}\cdot\mathbf{r}}\tilde{\sigma
}_{41}),\\
&  \frac{\partial\sigma_{22}}{\partial t}=\Gamma_{23}\sigma_{33}+\Gamma
_{24}\sigma_{44}+i(\Omega_{P2}^{\ast}e^{-i\mathbf{k}_{P2}\cdot\mathbf{r}%
}\tilde{\sigma}_{24}+g_{p}\hat{\mathcal{E}}_{p}^{\dagger}e^{-i\mathbf{k}%
_{p}\cdot\mathbf{r}}\tilde{\sigma}_{23}-\Omega_{P2}e^{i\mathbf{k}_{P2}%
\cdot\mathbf{r}}\tilde{\sigma}_{42}\nonumber\\
&  -g_{p}\hat{\mathcal{E}}_{p}e^{i\mathbf{k}_{p}\cdot\mathbf{r}}\tilde{\sigma
}_{32}),\\
&  \frac{\partial\sigma_{33}}{\partial t}=-\Gamma_{3}\sigma_{33}+i(\Omega
_{P1}e^{i\mathbf{k}_{P1}\cdot\mathbf{r}}\tilde{\sigma}_{31}+g_{p}%
\hat{\mathcal{E}}_{p}e^{i\mathbf{k}_{p}\cdot\mathbf{r}}\tilde{\sigma}%
_{32}-\Omega_{P1}^{\ast}e^{-i\mathbf{k}_{P1}\cdot\mathbf{r}}\tilde{\sigma
}_{13}\nonumber\\
&  -g_{p}\hat{\mathcal{E}}_{p}^{\dagger}e^{-i\mathbf{k}_{p}\cdot\mathbf{r}%
}\tilde{\sigma}_{23}),\\
&  \frac{\partial\sigma_{44}}{\partial t}=-\Gamma_{4}\sigma_{44}+i(\Omega
_{P2}e^{i\mathbf{k}_{P2}\cdot\mathbf{r}}\tilde{\sigma}_{42}+g_{c}%
\hat{\mathcal{E}}_{c}e^{i\mathbf{k}_{c}\cdot\mathbf{r}}\tilde{\sigma}%
_{41}-\Omega_{P2}^{\ast}e^{-i\mathbf{k}_{P2}\cdot\mathbf{r}}\tilde{\sigma
}_{24}\nonumber\\
&  -g_{c}\hat{\mathcal{E}}_{c}^{\dagger}e^{-i\mathbf{k}_{c}\cdot\mathbf{r}%
}\tilde{\sigma}_{14}),
\end{align}
where $\Gamma_{mn}$ is the population decay rate from the level $n$ to level
$m$, and we introduce slowly varying matrix elements in time: $\sigma
_{13}=\tilde{\sigma}_{13}e^{-i\omega_{p1}t}$, $\sigma_{14}=\tilde{\sigma}%
_{14}e^{-i\omega_{c}t}$, $\sigma_{24}=\tilde{\sigma}_{24}e^{-i\omega_{p2}t}$,
and $\sigma_{23}=\tilde{\sigma}_{23}e^{-i\omega_{p}t}$. In addition, the set
of equations for $\tilde{\sigma}_{nm}$ ($n\neq m$) are given by%
\begin{align}
\frac{\partial\tilde{\sigma}_{42}}{\partial t}  &  =[i(\delta-\Delta
_{2})-\gamma_{42}]\tilde{\sigma}_{42}-i[\Omega_{P2}^{\ast}e^{-i\mathbf{k}%
_{P2}\cdot\mathbf{r}}\sigma_{22,44}-g_{p}\hat{\mathcal{E}}_{p}^{\dagger
}e^{-i\mathbf{k}_{p}\cdot\mathbf{r}}\tilde{\sigma}_{43}+g_{c}\hat{\mathcal{E}%
}_{c}^{\dagger}e^{-i\mathbf{k}_{c}\cdot\mathbf{r}}\tilde{\sigma}%
_{12}],\nonumber\\
& \\
\frac{\partial\tilde{\sigma}_{41}}{\partial t}  &  =(-i\Delta_{2}-\gamma
_{41})\tilde{\sigma}_{41}+i[\Omega_{P1}^{\ast}e^{-i\mathbf{k}_{P1}%
\cdot\mathbf{r}}\tilde{\sigma}_{43}-g_{c}\hat{\mathcal{E}}_{c}^{\dagger
}e^{-i\mathbf{k}_{c}\cdot\mathbf{r}}\sigma_{11,44}-\Omega_{P2}^{\ast
}e^{-i\mathbf{k}_{P2}\cdot\mathbf{r}}\tilde{\sigma}_{12}],\\
\frac{\partial\tilde{\sigma}_{32}}{\partial t}  &  =[i(\delta-\Delta
_{1})-\gamma_{32}]\tilde{\sigma}_{32}+i[\Omega_{P2}^{\ast}e^{-i\mathbf{k}%
_{P2}\cdot\mathbf{r}}\tilde{\sigma}_{34}-g_{p}\hat{\mathcal{E}}_{p}^{\dagger
}e^{-i\mathbf{k}_{p}\cdot\mathbf{r}}\sigma_{22,33}-\Omega_{P1}^{\ast
}e^{-i\mathbf{k}_{P1}\cdot\mathbf{r}}\tilde{\sigma}_{12}],\nonumber\\
& \\
\frac{\partial\tilde{\sigma}_{31}}{\partial t}  &  =-(i\Delta_{1}+\gamma
_{31})\tilde{\sigma}_{31}-i[\Omega_{P1}^{\ast}e^{-i\mathbf{k}_{P1}%
\cdot\mathbf{r}}\sigma_{11,33}-g_{c}\hat{\mathcal{E}}_{c}^{\dagger
}e^{-i\mathbf{k}_{c}\cdot\mathbf{r}}\tilde{\sigma}_{34}+g_{p}\hat{\mathcal{E}%
}_{p}^{\dagger}e^{-i\mathbf{k}_{p}\cdot\mathbf{r}}\tilde{\sigma}_{21}],\\
\frac{\partial\tilde{\sigma}_{43}}{\partial t}  &  =[i(\Delta_{1}-\Delta
_{2})-\gamma_{43}]\tilde{\sigma}_{43}+i[\Omega_{P1}e^{i\mathbf{k}_{P1}%
\cdot\mathbf{r}}\tilde{\sigma}_{41}+g_{p}\hat{\mathcal{E}}_{p}e^{i\mathbf{k}%
_{p}\cdot\mathbf{r}}\tilde{\sigma}_{42}-\Omega_{P2}^{\ast}e^{-i\mathbf{k}%
_{P2}\cdot\mathbf{r}}\tilde{\sigma}_{23}\nonumber\\
&  -g_{c}\hat{\mathcal{E}}_{c}^{\dagger}e^{-i\mathbf{k}_{c}\cdot\mathbf{r}%
}\tilde{\sigma}_{13}],\\
\frac{\partial\tilde{\sigma}_{21}}{\partial t}  &  =-(i\delta+\gamma
_{21})\tilde{\sigma}_{21}+i[\Omega_{P1}^{\ast}e^{-i\mathbf{k}_{P1}%
\cdot\mathbf{r}}\tilde{\sigma}_{23}+g_{c}\hat{\mathcal{E}}_{c}^{\dagger
}e^{-i\mathbf{k}_{c}\cdot\mathbf{r}}\tilde{\sigma}_{24}-\Omega_{P2}%
e^{i\mathbf{k}_{P2}\cdot\mathbf{r}}\tilde{\sigma}_{41}\nonumber\\
&  -g_{p}\hat{\mathcal{E}}_{p}e^{i\mathbf{k}_{p}\cdot\mathbf{r}}\tilde{\sigma
}_{31}],
\end{align}
where $\sigma_{22,44}=\sigma_{22}-\sigma_{44}$, $\sigma_{11,44}=\sigma
_{11}-\sigma_{44}$, $\sigma_{22,33}=\sigma_{22}-\sigma_{33}$, and
$\sigma_{11,33}=\sigma_{11}-\sigma_{33}$, the single-photon detunings are
$\Delta_{1}=\omega_{P1}-\omega_{31}$ and $\Delta_{2}=\omega_{P2}-\omega
_{42}+\delta$, the two-photon detuning is $\delta=\omega_{P1}-\omega
_{p}-\omega_{21}$, and the slowly varying matrix elements are $\sigma
_{43}=\tilde{\sigma}_{43}e^{i(\omega_{p2}-\omega_{p})t}$ and $\sigma
_{12}=\tilde{\sigma}_{12}e^{-i(\omega_{c}-\omega_{p2})t}$. $\gamma_{nm}$ gives
the dephasing rate of the $\sigma_{nm}$ coherence, and $\gamma_{nm}%
=(\Gamma_{m}+\Gamma_{n})/2+\gamma_{nm}^{c}$, where $\Gamma_{n}$ is the total
decay rate out of level $n$ and $\gamma_{nm}^{c}$ is the dephasing rate due to
any other source of decoherence.

Now we are in a position to solve the properties of the system. For
convenience, we let $\Gamma_{3}=\Gamma_{4}\equiv\gamma$ and $\Gamma
_{13}=\Gamma_{14}=\Gamma_{23}=\Gamma_{24}=\gamma/2$, and the complex decay
rates are%
\begin{align}
\xi_{42}  &  =-\gamma/2+i(\delta-\Delta_{2}),\text{ }\xi_{41}=-\left(
\gamma/2+i\Delta_{2}\right)  ,\\
\xi_{32}  &  =-\gamma/2+i(\delta-\Delta_{1}),\text{ }\xi_{31}=-\left(
\gamma/2+i\Delta_{1}\right)  ,\\
\xi_{43}  &  =-\gamma+i(\Delta_{1}-\Delta_{2}),\text{ }\xi_{21}=-\left(
\gamma_{21}+i\delta\right)  .
\end{align}
In order to obtain analytical expressions, we assume that the pump fields
propagate without depletion, and the steady-state expectation values for the
zeroth-order atomic operators $\sigma_{33}$ and $\sigma_{44}$\ are equal to%
\begin{equation}
\langle\sigma_{33}\rangle=\langle\sigma_{44}\rangle=\frac{\left\vert
\Omega_{P1}\right\vert ^{2}\left\vert \Omega_{P2}\right\vert ^{2}}{\left\vert
\Omega_{P2}\right\vert ^{2}\left\vert \xi_{31}\right\vert ^{2}+\left\vert
\Omega_{P1}\right\vert ^{2}\left\vert \xi_{42}\right\vert ^{2}+4\left\vert
\Omega_{P1}\right\vert ^{2}\left\vert \Omega_{P2}\right\vert ^{2}}.
\end{equation}
Then, the population differences are given by%
\begin{align}
\langle\sigma_{11,33}\rangle &  =\langle\sigma_{11,44}\rangle=\frac{\left\vert
\xi_{31}\right\vert ^{2}\left\vert \Omega_{P2}\right\vert ^{2}}{\left\vert
\Omega_{P2}\right\vert ^{2}\left\vert \xi_{31}\right\vert ^{2}+\left\vert
\Omega_{P1}\right\vert ^{2}\left\vert \xi_{42}\right\vert ^{2}+4\left\vert
\Omega_{P1}\right\vert ^{2}\left\vert \Omega_{P2}\right\vert ^{2}},\\
\langle\sigma_{22,33}\rangle &  =\langle\sigma_{22,44}\rangle=\frac{\left\vert
\xi_{42}|^{2}|\Omega_{P1}\right\vert ^{2}}{\left\vert \xi_{31}\right\vert
^{2}\left\vert \Omega_{P2}\right\vert ^{2}+\left\vert \xi_{42}\right\vert
^{2}\left\vert \Omega_{P1}\right\vert ^{2}+4\left\vert \Omega_{P1}\Omega
_{P2}\right\vert ^{2}},
\end{align}
%%%
%%%
We also assume that the probe and conjugate fields are weak fields, such that
we only keep terms to first order in $\Omega_{p}$ and $\Omega_{c}$. The
steady-state density matrix elements $\tilde{\sigma}_{23}$ and $\tilde{\sigma
}_{14}$ are given by%
\begin{align}
\tilde{\sigma}_{23}  &  =\frac{i\xi_{41}}{D}\times\{g_{p}\hat{\mathcal{E}}%
_{p}e^{i\mathbf{k}_{p}\cdot\mathbf{r}}[\sigma_{11,33}\left(  \frac{\xi_{43}%
}{\xi_{31}}+\frac{\left\vert \Omega_{P1P2}\right\vert ^{2}}{\xi_{31}\xi_{41}%
}\right)  \left\vert \Omega_{P1}\right\vert ^{2}-(\frac{\xi_{43}\left\vert
\Omega_{P2}\right\vert ^{2}+\xi_{21}\left\vert \Omega_{P1}\right\vert ^{2}%
}{\xi_{41}}\nonumber\\
&  +\xi_{21}\xi_{43})\sigma_{22,33}+\left\vert \Omega_{P2}\right\vert
^{2}\left(  \frac{\xi_{21}}{\xi_{42}}-\frac{\left\vert \Omega_{P1P2}%
\right\vert ^{2}}{\xi_{42}\xi_{41}}\right)  \sigma_{22,44}]+\Omega_{P1}%
\Omega_{P2}[\frac{\xi_{21}+\xi_{43}}{\xi_{41}}\sigma_{11,44}\nonumber\\
&  +\left(  \frac{\xi_{43}}{\xi_{42}^{\ast}}+\frac{\left\vert \Omega
_{P1P2}\right\vert ^{2}}{\xi_{42}^{\ast}\xi_{41}}\right)  \sigma
_{22,44}+\left(  \frac{\xi_{21}}{\xi_{31}^{\ast}}-\frac{\left\vert
\Omega_{P1P2}\right\vert ^{2}}{\xi_{31}^{\ast}\xi_{41}}\right)  \sigma
_{11,33}]e^{i\Delta\mathbf{k}_{p}\cdot\mathbf{r}}g_{c}\hat{\mathcal{E}}%
_{c}^{\dag}\},
\end{align}%
\begin{align}
&  \tilde{\sigma}_{14}=\frac{i\xi_{32}}{D^{\ast}}\times\{g_{c}\hat
{\mathcal{E}}_{c}e^{i\mathbf{k}_{c}\cdot\mathbf{r}}[\sigma_{11,33}\left(
\frac{\xi_{21}^{\ast}}{\xi_{31}}+\frac{\left\vert \Omega_{P1P2}\right\vert
^{2}}{\xi_{31}\xi_{32}}\right)  \left\vert \Omega_{P1}\right\vert ^{2}%
-(\frac{\xi_{21}^{\ast}\left\vert \Omega_{P2}\right\vert ^{2}+\xi_{43}^{\ast
}\left\vert \Omega_{P1}\right\vert ^{2}}{\xi_{32}}\nonumber\\
&  +\xi_{21}^{\ast}\xi_{43}^{\ast})\sigma_{11,44}+\left\vert \Omega
_{P2}\right\vert ^{2}\left(  \frac{\xi_{43}^{\ast}}{\xi_{42}}-\frac{\left\vert
\Omega_{P1P2}\right\vert ^{2}}{\xi_{42}\xi_{32}}\right)  \sigma_{22,44}%
]+\Omega_{P1}\Omega_{P2}[\frac{\xi_{43}^{\ast}+\xi_{21}^{\ast}}{\xi_{32}%
}\sigma_{22,33}\nonumber\\
&  +\left(  \frac{\xi_{21}^{\ast}}{\xi_{42}^{\ast}}+\frac{\left\vert
\Omega_{P1P2}\right\vert ^{2}}{\xi_{42}^{\ast}\xi_{32}}\right)  \sigma
_{22,44}+\left(  \frac{\xi_{43}^{\ast}}{\xi_{31}^{\ast}}-\frac{\left\vert
\Omega_{P1P2}\right\vert ^{2}}{\xi_{31}^{\ast}\xi_{32}}\right)  \sigma
_{11,33}]e^{i\Delta\mathbf{k}_{c}\cdot\mathbf{r}}g_{p}\hat{\mathcal{E}}%
_{p}^{\dag}\},
\end{align}
where
\begin{align}
&  D=\xi_{41}\xi_{21}\xi_{43}\xi_{32}^{\ast}+(\xi_{41}\xi_{43}+\xi_{21}%
\xi_{32}^{\ast})\left\vert \Omega_{P1}\right\vert ^{2}+(\xi_{41}\xi_{21}%
+\xi_{43}\xi_{32}^{\ast})\left\vert \Omega_{P2}\right\vert ^{2}+\left\vert
\Omega_{P1P2}\right\vert ^{4},\nonumber\\
&  \left\vert \Omega_{P1P2}\right\vert ^{2}=\left\vert \Omega_{P1}\right\vert
^{2}-\left\vert \Omega_{P2}\right\vert ^{2},\Delta\mathbf{k}_{p}%
=\mathbf{k}_{P1}+\mathbf{k}_{P2}-\mathbf{k}_{c},\Delta\mathbf{k}%
_{c}=\mathbf{k}_{P1}+\mathbf{k}_{P2}-\mathbf{k}_{p}.
\end{align}

Using the atomic operators to evaluate the linear and nonlinear components of
the polarization at $\omega_{p}$ and $\omega_{c}$, the polarization of the
atomic medium at a particular frequency is given by $\hat{P}\left(  \omega
_{p}\right)  =Nd_{23}\tilde{\sigma}_{23}+H.c.$ and $\hat{P}\left(  \omega
_{c}\right)  =Nd_{14}\tilde{\sigma}_{14}+H.c.$. The polarization of the medium
at frequency $\omega_{i}$ can be divided into two different terms: one that is
proportional to the field at frequency $\omega_{i}$ and one that is
proportional to the field at frequency $\omega_{P1}+\omega_{P2}-\omega_{i}$,
such that
\begin{align}
\hat{P}_{p}(\omega_{p})  &  =\sqrt{\frac{\hbar\omega_{p}}{2\epsilon_{0}V}%
}\epsilon_{0}\chi_{pp}(\omega_{p})\hat{\mathcal{E}}_{p}e^{i\mathbf{k}_{p}%
\cdot\mathbf{r}}+\sqrt{\frac{\hbar\omega_{c}}{2\epsilon_{0}V}}\epsilon_{0}%
\chi_{pc}(\omega_{p})\hat{\mathcal{E}}_{c}^{\dag}e^{i\Delta\mathbf{k}_{p}%
\cdot\mathbf{r}}+\mathrm{H.c.},\\
\hat{P}_{c}(\omega_{c})  &  =\sqrt{\frac{\hbar\omega_{c}}{2\epsilon_{0}V}%
}\epsilon_{0}\chi_{cc}(\omega_{c})\hat{\mathcal{E}}_{c}e^{i\mathbf{k}_{c}%
\cdot\mathbf{r}}+\sqrt{\frac{\hbar\omega_{p}}{2\epsilon_{0}V}}\epsilon_{0}%
\chi_{cp}(\omega_{c})\hat{\mathcal{E}}_{p}^{\dag}e^{i\Delta\mathbf{k}_{c}%
\cdot\mathbf{r}}+\mathrm{H.c.},
\end{align}
where the two coefficients $\chi_{pp}$, and $\chi_{cc}$ are given as follows:
\begin{align}
\chi_{pp}  &  =\frac{iN\left\vert d_{23}\right\vert ^{2}\xi_{41}}{\epsilon
_{0}\hbar D}[\left\vert \Omega_{P1}\right\vert ^{2}\left(  \frac{\xi_{43}}%
{\xi_{31}}+\frac{\left\vert \Omega_{P1P2}\right\vert ^{2}}{\xi_{31}\xi_{41}%
}\right)  \sigma_{11,33}-(\frac{\xi_{43}\left\vert \Omega_{P2}\right\vert
^{2}+\xi_{21}\left\vert \Omega_{P1}\right\vert ^{2}}{\xi_{41}}\nonumber\\
&  +\xi_{21}\xi_{43})\sigma_{22,33}+\left\vert \Omega_{P2}\right\vert
^{2}\left(  \frac{\xi_{21}}{\xi_{42}}-\frac{\left\vert \Omega_{P1P2}%
\right\vert ^{2}}{\xi_{42}\xi_{41}}\right)  \sigma_{22,44}],
\end{align}%
\begin{align}
\chi_{cc}  &  =\frac{iN\left\vert d_{14}\right\vert ^{2}\xi_{32}}{\epsilon
_{0}\hbar D^{\ast}}[\left\vert \Omega_{P1}\right\vert ^{2}\left(  \frac
{\xi_{21}^{\ast}}{\xi_{31}}+\frac{\left\vert \Omega_{P1P2}\right\vert ^{2}%
}{\xi_{31}\xi_{32}}\right)  \sigma_{11,33}-(\frac{\xi_{21}^{\ast}\left\vert
\Omega_{P2}\right\vert ^{2}+\xi_{43}^{\ast}\left\vert \Omega_{P1}\right\vert
^{2}}{\xi_{32}}\nonumber\\
&  +\xi_{21}^{\ast}\xi_{43}^{\ast})\sigma_{11,44}+\left\vert \Omega
_{P2}\right\vert ^{2}\left(  \frac{\xi_{43}^{\ast}}{\xi_{42}}-\frac{\left\vert
\Omega_{P1P2}\right\vert ^{2}}{\xi_{42}\xi_{32}}\right)  \sigma_{22,44}],
\end{align}
The two coefficients $\chi_{pc}$ and $\chi_{cp}$ are given by%
\begin{align}
\chi_{pc}  &  =\frac{iNd_{23}d_{14}\xi_{41}\Omega_{P1}\Omega_{P2}}%
{\varepsilon_{0}\hbar D}[\frac{\xi_{21}+\xi_{43}}{\xi_{41}}\sigma
_{11,44}+(\frac{\xi_{43}}{\xi_{42}^{\ast}}+\frac{\left\vert \Omega
_{P1P2}\right\vert ^{2}}{\xi_{42}^{\ast}\xi_{41}})\sigma_{22,44}\nonumber\\
&  +\left(  \frac{\xi_{21}}{\xi_{31}^{\ast}}-\frac{\left\vert \Omega
_{P1P2}\right\vert ^{2}}{\xi_{31}^{\ast}\xi_{41}}\right)  \sigma_{11,33}],
\end{align}%
\begin{align}
\chi_{cp}  &  =\frac{iNd_{23}d_{14}\xi_{32}\Omega_{P1}\Omega_{P2}}%
{\varepsilon_{0}\hbar D^{\ast}}[\frac{\xi_{43}^{\ast}+\xi_{21}^{\ast}}%
{\xi_{32}}\sigma_{22,33}+(\frac{\xi_{21}^{\ast}}{\xi_{42}^{\ast}}%
+\frac{\left\vert \Omega_{P1P2}\right\vert ^{2}}{\xi_{42}^{\ast}\xi_{32}%
})\sigma_{22,44}\nonumber\\
&  +\left(  \frac{\xi_{43}^{\ast}}{\xi_{31}^{\ast}}-\frac{\left\vert
\Omega_{P1P2}\right\vert ^{2}}{\xi_{31}^{\ast}\xi_{32}}\right)  \sigma
_{11,33}].
\end{align}
The two coefficients $\chi_{pp}$ and $\chi_{cc}$ describe the effective linear
polarization processes for the probe and conjugate fields, respectively, and
unlike the usual linear coefficients, they depend nonlinearly on the pump
field. The other two coefficients $\chi_{pc}$ and $\chi_{cp}$ are responsible
for the 4WM process.

Since the two pump fields have the same polarizations and frequencies, the two
pump fields $E_{P1}$ and $E_{P2}$\ also couple the transitions $|2\rangle
\rightarrow|4\rangle$ and $|1\rangle\rightarrow|3\rangle$ with probabilities
of $\frac{1}{2}$. Due to coupling another set of transitions with the same
effective electric dipole, we just swap the Rabi frequencies $\Omega_{P1}$ and
$\Omega_{P2}$ in the above 4 coefficients $\chi_{pp}$, $\chi_{cc}$, $\chi
_{pc}$ and $\chi_{cp}$ to obtain a new set of transitions.


\begin{thebibliography}{99}                                                                                               %


\bibitem {Simon10}C. Simon, \emph{et al.} \textquotedblleft Quantum memories.
A review based on the European integrated project \textquotedblleft Qubit
Applications (QAP)\textquotedblright." Eur. Phys. J. D \textbf{58}, 1 (2010).

\bibitem {Tanimura}T. Tanimura, D. Akamatsu, Y. Yokoi, A. Furusawa, and M.
Kozuma, \textquotedblleft Generation of squeezed vacuum resonant on a rubidium
D1 line with periodically poled KTiOPO4,\textquotedblright\ Opt. Lett. 31,
2344-2346 (2006).

\bibitem {Hetet}G. Hetet, O. Glockl, K. A. Pilypas, C. C. Harb, B.C. Buchler,
H. A. Bachor, and P. K. Lam, \textquotedblleft Squeezed light for
bandwith-limited atom optics experiments at the rubidium D1
line,\textquotedblright\ J. Phys. B: At. Mol. Opt. Phys. 40, 221-226 (2007).

\bibitem {Polzik}E. S. Polzik, J. Carri, and H. J. Kimble,\textquotedblleft
Spectroscopy with squeezed light,\textquotedblright\ Phys. Rev. Lett. 68, 3020 (1992).

\bibitem {Marin}F. Marin, A. Bramati, V. Jost, and E.
Giacobino,\textquotedblleft Demonstration of high sensitivity spectroscopy
with squeezed semiconductor lasers,\textquotedblright\ Optics Commun. 140, 146 (1997).

\bibitem {Burks09}S. Burks, J. Ortalo, A. Chiummo, X. Jia, F. Villa, A.
Bramati, J. Laurat, and E. Giacobino, "Vacuum squeezed light for atomic
memories at the D$_{2}$ cesium line", Opt. Express \textbf{17}, 3777 (2009).

\bibitem {Honda}K. Honda, D. Akamatsu, M. Arikawa, Y. Yokoi, K. Akiba, S.
Nagatsuka, T. Tanimura, A. Furusawa, and M. Kozuma, \textquotedblleft Storage
and Retrieval of a Squeezed Vacuum,\textquotedblright\ Phys. Rev. Lett.
100,093601 (2008).

\bibitem {Appel}J. Appel, E. Figueroa, D. Korystov, M. Lobino, and A. I.
Lvovsky, \textquotedblleft Quantum memory for squeezed
light,\textquotedblright\ Phys. Rev. Lett. 100, 093602 (2008).

\bibitem {Guo19}J. Guo, X. Feng, P. Yang, Z. Yu, L. Q. Chen, C.-H. Yuan, W.
Zhang, \textquotedblleft High-performance Raman quantum memory with optimal
control," Nature Communications \textbf{10}, 148 (2018).

\bibitem {Slusher}R. E. Slusher, L. W. Hollberg, B. Yurke, J. C. Mertz, and J.
F. Valley, ``Observation of squeezed states generated by four-wave mixing in
an optical cavity," Phys. Rev. Lett. \textbf{55}, 2409 (1985).

\bibitem {Schnabel}R. Schnabel, ``Squeezed states of light and their
applications in laser interferometers," Phys. Rep. \textbf{684}, 1 (2017).

\bibitem {Hemmer}P. R. Hemmer, D. P. Katz, J. Donoghue, M. Cronin-Golomb, M.
S. Shariar, and P. Kumar, ``Efficient low-intensity optical phase conjugation
based on coherent population trapping in sodium," Opt. Lett. \textbf{20}, 982 (1995).

\bibitem {Lukin}M. D. Lukin, P. R. Hemmer, M. L\"{o}ffler, and M. O. Scully,
``Resonant enhancement of parametric processes via radiative interference and
induced coherence," Phys. Rev. Lett. \textbf{81}, 2675 (1998).

\bibitem {Zibrov}A. S. Zibrov, M. D. Lukin, and M. O. Scully, ``Nondegenerate
parametric self-oscillation via multiwave mixing in coherent atomic media,"
Phys. Rev. Lett. \textbf{83}, 4049 (1999).

\bibitem {Balic}V. Balic, D. A. Braje, P. Kolchin, G. Y. Yin, and S. E.
Harris, ``Generation of paired photons with controllable waveforms," Phys.
Rev. Lett. \textbf{94}, 183601 (2005).

\bibitem {Kolchin}P. Kolchin, S. Du, C. Belthangady, G. Y. Yin, and S. E.
Harris, ``Generation of narrow-bandwidth paired photons: use of a single
driving laser," Phys. Rev. Lett. \textbf{97}, 113602 (2006).

\bibitem {Thompson}J. K. Thompson, J. Simon, H. Loh, and V. Vuletic, ``A
high-brightness source of narrowband, identical-photon pairs," Science
\textbf{313}, 74 (2006).
%%%%


\bibitem {Wal}C. H. van der Wal, M. D. Eisaman, A. Andr\'{e}, R. L. Walsworth,
D. F. Phillips, A. S. Zibrov, and M. D. Lukin, ``Atomic memory for correlated
photon states," Science \textbf{301}, 196 (2003).

\bibitem {McCormick07}C. F. McCormick, V. Boyer, E. Arimondo, and P. D. Lett,
``Strong relative intensity squeezing by four-wave mixing in rubidium vapor,"
Opt. Lett. \textbf{32}, 178 (2007).

\bibitem {McCormick08}C. F. McCormick, A. M. Marino, V. Boyer, and P. D. Lett,
``Strong low-frequency quantum correlations from a four-wave-mixing
amplifier," Phys. Rev. A. \textbf{78}, 043816 (2008).

\bibitem {Zhang15}Z. Zhang, F. Wen, J. Che, D. Zhang, C. Li, Y. Zhang, and M.
Xiao, ``Dressed gain from the parametrically amplified four-wave mixing
process in an atomic vapor," Sci. Rep. 5, 15058 (2015).

\bibitem {Zhang17}D. Zhang, C. Li, Z. Zhang, Y. Zhang, Y. Zhang, and M. Xiao,
``Enhanced intensity-difference squeezing via energy-level modulations in hot
atomic media," Phys. Rev. A 96, 043847 (2017).

\bibitem {Liu11}C. Liu, J. Jing, Z. Zhou, R. C. Pooser, F. Hudelist, L. Zhou,
and W. Zhang, ``Realization of low frequency and controllable bandwidth
squeezing based on a four-wave-mixing amplifier in rubidium vapor," Opt. Lett.
\textbf{36}, 2979 (2011).

\bibitem {Ma18}R. Ma, W. Liu, Z. Qin, X. Su, X. Jia, J. Zhang, and J. Gao,
``Compact sub-kilohertz low-frequency quantum light source based on four-wave
mixing in cesium vapor," Opt. Lett. \textbf{43}, 1243 (2018).

\bibitem {Wu19}M.-C. Wu, B. L. Schmittberger, N. R. Brewer, R. W. Speirs, K.
M. Jones, P. D. Lett, ``Twin-beam intensity-difference squeezing below 10 Hz,"
Opt. Express \textbf{27}, 4769 (2019).

\bibitem {Boyer08}V. Boyer, A. M. Marino, and P. D. Lett, ``Generation of
spatially broadband twin beams for quantum imaging," Phys. Rev. Lett.
\textbf{100}, 143601 (2008).

\bibitem {Boyer082}V. Boyer, A. M. Marino, R. C. Pooser, and P. D. Lett,
``Entangled images from four-wave mixing," Science, \textbf{321}, 544 (2008).

\bibitem {Qin14}Z. Qin, L. Cao, H. Wang, A. M. Marino, W. Zhang, and J. Jing,
``Experimental generation of multiple quantum correlated beams from hot
rubidium Vapor," Phys. Rev. Lett. 113, 023602 (2014).

\bibitem {Qin15}Z. Qin, L. Cao, and J. Jing, ``Experimental characterization
of quantum correlated triple beams generated by cascaded four-wave mixing
processes," Appl. Phys. Lett. 106, 211104 (2015).

\bibitem {Hudelist}F. Hudelist, J. Kong, C. Liu, J. Jing, Z. Y. Ou, and W.
Zhang, ``Quantum metrology with parametric amplifier-based photon correlation
interferometers," Nat. Commun. \textbf{5}, 3049 (2014).

\bibitem {Du18}W. Du, J. Jia, J. F. Chen, Z. Y. Ou, and W. Zhang, ``Absolute
sensitivity of phase measurement in an SU(1,1) type interferometer," Opt.
Lett. \textbf{43}, 1051 (2018).

\bibitem {Corzo12}N. V. Corzo, A. M. Marino, K. M. Jones, and P. D. Lett,
``Noiseless optical amplifier operating on hundreds of spatial modes," Phys.
Rev. Lett. \textbf{109}, 043602 (2012).

\bibitem {Embrey15}C. S. Embrey, M. T. Turnbull, P. G. Petrov, and V. Boyer,
``Observation of localized multi-spatial-mode quadrature squeezing," Phys.
Rev. X \textbf{5}, 031004 (2015)

\bibitem {Wang17}L. H. Wang, C. Fabre, and J. Jing, ``Single-step fabrication
of scalable multimode quantum resources using four-wave mixing with a
spatially structured pump," Phys. Rev. A 95, 051802 (2017).

\bibitem {Cao17}L. Cao, J. Qi, J. Du, and J. Jing, ``Experimental generation
of quadruple quantum-correlated beams from hot rubidium vapor by cascaded
four-wave mixing using spatial multiplexing," Phys. Rev. A 95, 023803 (2017).

\bibitem {Jia}J. Jia, W. Du, J. F. Chen, C.-H. Yuan, Z. Y. Ou, and W. Zhang,
\textquotedblleft Generation of frequency degenerate twin beams in 85Rb
vapor," Opt. Lett. \textbf{42}, 4024 (2017).

\bibitem {Knutson18}E. M. Knutson, J. D. Swaim, S. Wyllie, and R. T. Glasser,
\textquotedblleft Optimal mode configuration for multiple phase-matched
four-wave-mixing processes," Phys. Rev. A \textbf{98}, 013828 (2018).

\bibitem {Kauranen}M. Kauranen, J. J. Maki, A. L. Gaeta, and R. W. Boyd,
\textquotedblleft Two-beam-excited conical emission," Opt. Lett. \textbf{16},
943 (1991).

\bibitem {Fox}R. W. Fox, \textquotedblleft Trace detection with diode lasers,"
Ph. D. Thesis, University of Colorado, Boulder (1995).

\bibitem {Steck}D. A. Steck, \textquotedblleft Rubidium 85 D Line Data," http://steck.us/alkalidata

\bibitem {Corzo11}N. Corzo, A. M. Marino, K. M. Jones, and P. D. Lett,
\textquotedblleft Multi-spatial-mode single-beam quadrature squeezed states of
light from four-wave mixing in hot rubidium vapor," Opt. Express, \textbf{19},
21358 (2011).

\bibitem {Davis}W. V. Davis, M. Kauranen, E. M. Nagasako, R. J. Gehr, A. L.
Gaeta, and R. W. Boyd, \textquotedblleft Excess noise acquired by a laser beam
after propagating through an atomic-potassium vapor,\textquotedblright\ Phys.
Rev. A 51, 4152--4159 (1995).

\bibitem {Jasperse11}M. Jasperse, L. D. Turner, and R. E. Scholten,
\textquotedblleft Relative intensity squeezing by four-wave mixing with loss:
an analytic model and experimental diagnostic," Opt. Express \textbf{19}, 3765 (2011).

\bibitem {Turnbull13}M. T. Turnbull, P. G. Petrov, C. S. Embrey, A. M. Marino,
and V. Boyer, \textquotedblleft Role of the phase-matching condition in
nondegenerate four-wave mixing in hot vapors for the generation of squeezed
states of light," Phys. Rev. A. \textbf{88}, 033845 (2013).

\bibitem {Boyer07}V. Boyer, C. F. McCormick, E. Arimondo, and P. D. Lett,
``Ultraslow Propagation of matched pulses by four-wave mixing in an atomic
vapor," Phys. Rev. Lett. 99, 143601 (2007).

\bibitem {Glasser12}R. T. Glasser, Ulrich Vogl, and Paul D. Lett, ``Stimulated
Generation of superluminal light pulses via four-wave mixing," Phys. Rev.
Lett. 108, 173902 (2012).
\end{thebibliography}
\end{document}